\documentclass[12pt]{article}
\usepackage{times}

\usepackage{epsf}
\def\beq{\begin{equation}}
\def\eeq{\end{equation}}
\def\bea{\begin{eqnarray}}
\def\eea{\end{eqnarray}}

\def\vel{\left|}
\def\ver{\right|}
\def\nnb{\nonumber}
\def\ga{\left(}
\def\dr{\right)}
\def\aga{\left\{}
\def\adr{\right\}}

\def\nnb{\nonumber}
\def\la{\langle}
\def\ra{\rangle}
\def\ba{\begin{array}}
\def\ea{\end{array}}
\def\bea{\begin{eqnarray}}
\def\eea{\end{eqnarray}}

\title{ {\bf
$B\rightarrow K^* \tau^+ \tau^-$  decay in the general two Higgs doublet 
model including the neutral Higgs boson effects }}
\author{\vspace{1cm}\\
         {\bf E. O. Iltan} 
         \thanks{E-mail address:
        eiltan@heraklit.physics.metu.edu.tr},\,\,\,\,
        {\bf G. Turan}
        \thanks{E-mail address:
        gsevgur@rorqual.metu.edu.tr}\,\,  and \,\,\,
       {\bf I. Turan}
        \thanks{E-mail address:
        Ituran@metu.edu.tr}
 \\
        Physics Department, Middle East Technical University \\
        Ankara, Turkey\\}

\date{}

\begin{document}
\setlength{\baselineskip}{24pt}
\maketitle
\setlength{\baselineskip}{7mm}
\begin{abstract}
We study the CP violating asymmetry, the forward-backward asymmetry of the 
lepton pair and the CP asymmetry in the forward backward asymmetry for the
exclusive decay $B\rightarrow K^* \tau^+ \tau^-$ in the general two 
Higgs doublet model including the neutral Higgs boson effects. 
We analyse the dependencies of these quantities on the model III parameters, 
We found that the physical parameters studied above are at the order of the 
magnitude $1\%$ and neutral Higgs boson effects are detectable for large 
values of the coupling $\bar{\xi}_{N,\tau\tau}^D$.
\end{abstract} 
\thispagestyle{empty}
\newpage
\setcounter{page}{1}
\section{Introduction}
Rare B decays are induced by flavor changing neutral currents (FCNC) at loop 
level in the Standard model (SM) and they are rich phenomenologically. They 
open a window for the determination of free parameters in the SM and also
the investigation of the physics beyond, such as, two Higgs Doublet 
model (2HDM), Minimal Supersymmetric extension of the SM (MSSM) \cite{Hewett}, 
etc.  The experimental work for rare B decays continue at  SLAC 
(BaBar), KEK (BELLE), B-Factories, DESY (HERA-B) and this stimulates the
theoretical effort on them.

The exclusive $B\rightarrow K^* l^+ l^-$ process is an important candidate 
among rare B decays. Since its branching ratio ($Br$) predicted in the SM is 
large, there is a strong hope that this physical quantity is measured in the
near future. The inclusive decay which induces $B\rightarrow K^* l^+ l^-$ 
process is $b\rightarrow s l^+l^-$ transition and it is extensively studied
in the literature, in the framework of the SM, 2HDM and MSSM. In 
\cite{Hou}-\cite{Jaus}, $b\rightarrow s l^+l^-$ process is studied for light 
lepton pairs, namely $l=e,\mu$.  In this case, the neutral Higgs boson (NHB) 
effects can be neglected since those contributions are proportional to the 
light lepton masses or corresponding Yukawa couplings. However for $l=\tau$, 
the NHB effects give sizable contributions. In \cite{Dai,Logan}  
$B\rightarrow X_s\tau^+\tau^-$ process was studied in the
model I, II versions the 2HDM and  it was shown that NHB effects are
important for large values of $tan\beta$. Currently the inclusive 
$b\rightarrow s l^+l^-$ decay was studied in the model III version of 2HDM 
\cite {Ergur} and it was observed that NHB effects can give considerable
contribution if the Yukawa interaction between $\tau$ lepton and neutral
Higgs bosons is large.

The theoretical analysis of exclusive decays is difficult due to the
hadronic form factors which contain uncertainities. However, their 
experimental investigation is easier compared to the one for the inclusive
decays. The calculation of physical observables in the hadronic level needs
non-perturbative methods. In the literature there are different studies
based on different approches, such as relativistic quark model by lightfront
formalism \cite{Jaus}, chiral theory \cite{Casalbuoni}, three point QCD sum
rules method \cite{Colangelo}, effective heavy quark theory \cite{Roberts}
and light cone QCD sum rules \cite{Alievsum,Pall}.

With the measured upper limit $5.2\times 10^{-6}$  ($4.0\times 10^{-6}$) 
for the $Br$ of the decay $B^+\rightarrow K^+ \mu^+\mu^-$ ($B^0\rightarrow 
K^{*0} \mu^+\mu^-$) \cite {Affolder}, the process $B\rightarrow K^* l^+l^-$ 
have reached great interest. There are various studies on these decays in 
the SM, SM with fourth generation, multi Higgs doublet models, MSSM and 
in a model independent way, in the literature \cite{Hou}-\cite{Jaus} and 
\cite{Casalbuoni}-\cite{AlievCakOzSav2}.  

The CP violating effect is an important physical quantity to ensure the
information about the free parameters of the model used. Since the CP
violation for $B\rightarrow K^* l^+l^-$ decay almost vanishes in the SM 
due to the unitarity of Cabibbo-Kobayashi-Maskawa (CKM) matrix elements 
and smallness of the term $V_{ub}\,V^*_{us}$, we have a chance to 
investigate the physics beyond the SM by searching the CP violating effects. 
In the model III version of the 2HDM, there is a new source for CP violation, 
namely complex Yukawa couplings. In \cite{Eril1,Eril2}, CP violating effects 
due to new phases in the model III and three Higgs doublet model, 
$3HDM(O_2)$, were studied and it was observed that a considerable CP 
asymmetry was obtained. 

In this work, we study the exclusive $B\rightarrow K^*\tau^+\tau^-$ decay 
in the general 2HDM , so-called model III, by including NHB effects. We use
the quark level effective Hamiltonian which is calculated in \cite{Ergur} 
and investigate the $A_{CP}$ and the forward-backward asymmetry
($A_{FB}$) of the lepton pair for the process underconsideration. Further, 
we calculate the CP asymmetry in $A_{FB}$, $(A_{CP}(A_{FB}))$ and observe 
that it can be measured in the forthcoming experiments. 

The paper is organized as follows:
In Section 2, we present the leading order (LO) QCD corrected effective
Hamiltonian  and the corresponding matrix element for the inclusive 
$b\rightarrow s\tau^+\tau^-$ decay, including NHB effects. Further, we give
the matrix element for the exclusive $B\rightarrow K^*\tau^+\tau^-$ decay
and the explicit expressions for $A_{FB}$, $A_{CP}$ and $A_{CP}(A_{FB})$. 
Section 3 is devoted to the analysis of the  dependencies of $A_{FB}$, 
$A_{CP}$ and $A_{CP}(A_{FB})$ on the CP parameter $sin\,\theta$, Yukawa 
coupling $\bar{\xi}_{N,\tau\tau}^{D}$ and the mass ratio $\frac{m_{h^0}}
{m_{A^0}}$ and to the discussion of our results. In Appendices, we give 
the explicit forms of the operators appearing in the effective Hamiltonian, 
the corresponding Wilson coefficients and the form factors existing in the 
hadronic matrix elements.

\section{\bf The exclusive $B\rightarrow K^* \tau^+ \tau^-$ decay in 
the model III including NHB effects. }
In the model III, the flavour changing neutral currents in the tree 
level are permitted and various new parameters, such as Yukawa couplings,
masses of new Higgs bosons, exist.  Yukawa couplings describe the 
interaction of fermions with gauge bosons. Our starting point is the 
inclusive $b\rightarrow s \tau^+ \tau^-$ process which induces the 
exclusive $B\rightarrow K^* \tau^+ \tau^-$ decay and the corresponding 
Yukawa interaction is given by
\begin{eqnarray}
{\cal{L}}_{Y}&=&\eta^{U}_{ij} \bar{Q}_{i L} \tilde{\phi_{1}} U_{j R}+
\eta^{D}_{ij} \bar{Q}_{i L} \phi_{1} D_{j R}+
\xi^{U\, \dagger}_{ij} \bar{Q}_{i L} \tilde{\phi_{2}} U_{j R}+
\xi^{D}_{ij} \bar{Q}_{i L} \phi_{2} D_{j R}\nonumber \\ &+& 
\eta^{D}_{k l} \bar{l}_{k L} \phi_{1} E_{l R}+
\xi^{D}_{k l} \bar{l}_{k L} \phi_{2} E_{l R}  
+ h.c. \,\,\, ,
\label{lagrangian}
\end{eqnarray}
where $i,j$ ($k,l$) are family indices of quarks (leptons), $L$ and $R$ 
denote chiral projections $L(R)=1/2(1\mp \gamma_5)$, $\phi_{m}$ for $m=1,2$, 
are the two scalar doublets, $Q_{i L}$ ($l_{k L}$) are quark (lepton) 
doublets, $U_{j R}$, $D_{j R}$ ($E_{l R}$) are the corresponding quark 
(lepton) singlets, $\eta^{U,D}_{ij}$ and $\xi^{U,D}_{ij}$ are the matrices 
of the Yukawa couplings which have complex entries in general. 
Here $\phi_{1}$ and $\phi_{2}$ are chosen as
\begin{eqnarray}
\phi_{1}=\frac{1}{\sqrt{2}}\left[\left(\begin{array}{c c} 
0\\v+H^{0}\end{array}\right)\; + \left(\begin{array}{c c} 
\sqrt{2} \chi^{+}\\ i \chi^{0}\end{array}\right) \right]\, ; 
\phi_{2}=\frac{1}{\sqrt{2}}\left(\begin{array}{c c} 
\sqrt{2} H^{+}\\ H_1+i H_2 \end{array}\right) \,\, .
\label{choice}
\end{eqnarray}
with the vacuum expectation values,  
\begin{eqnarray}
<\phi_{1}>=\frac{1}{\sqrt{2}}\left(\begin{array}{c c} 
0\\v\end{array}\right) \,  \, ; 
<\phi_{2}>=0 \,\, .
\label{choice2}
\end{eqnarray}
With this choice, the SM particles can be collected in the first doublet 
and the new particles in the second one. Further, we take $H_{1}$, $H_{2}$ 
as the mass eigenstates $h^{0}$, $A^{0}$ respectively. Note that, at tree 
level, there is no mixing among CP even neutral Higgs bosons, namely 
the SM one, $H^0$, and beyond, $h^{0}$.

The part which produce FCNC at tree level is  
\begin{eqnarray}
{\cal{L}}_{Y,FC}=
\xi^{U\,\dagger}_{ij} \bar{Q}_{i L} \tilde{\phi_{2}} U_{j R}+
\xi^{D}_{ij} \bar{Q}_{i L} \phi_{2} D_{j R} +
\xi^{D}_{kl} \bar{l}_{k L} \phi_{2} E_{l R} + h.c. \,\, .
\label{lagrangianFC}
\end{eqnarray}
In eq.(\ref{lagrangianFC}) the couplings  $\xi^{U,D}$ for the FC charged 
interactions are 
\begin{eqnarray}
\xi^{U}_{ch}&=& \xi_{neutral} \,\, V_{CKM} \nonumber \,\, ,\\
\xi^{D}_{ch}&=& V_{CKM} \,\, \xi_{neutral} \,\, ,
\label{ksi1} 
\end{eqnarray}
where  $\xi^{U,D}_{neutral}$ 
is defined by the expression
\begin{eqnarray}
\xi^{U (D)}_{N}=(V_{R(L)}^{U (D)})^{-1} \xi^{U,(D)} V_{L(R)}^{U (D)}
\,\, .
\label{ksineut}
\end{eqnarray}
and $\xi^{U,D}_{neutral}$ is denoted as $\xi^{U,D}_{N}$. 
Here the charged couplings are  the linear combinations of neutral 
couplings multiplied by $V_{CKM}$ matrix elements (see \cite{Alil1} for 
details). 

At this stage we would like to present the calculation of the matrix element 
for the inclusive $b\rightarrow s \tau^+ \tau^-$  decay, briefly. 
The procedure is the following:
\begin{itemize}
\item The calculation of the full theory including the NHB effects which 
comes from the interactions of neutral Higgs bosons $H^0$, $h^0$ and $A^0$ 
with $\tau$ lepton. 

\item Overcoming the logarithmic divergences by using the on-shell 
renormalization scheme. Here the renormalized vertex function is taken 
as
\begin{eqnarray}
\Gamma_{neutr}^{Ren}(p^2)=\Gamma_{neutr}^{0}(p^2)+\Gamma_{neutr}^{C} ,  
\label{vertren}
\end{eqnarray}
with the renormalization condition 
\begin{eqnarray}
\Gamma_{neutr}^{Ren}(p^2=m^2_{neutr})=0 ,
\label{rencond}
\end{eqnarray}
and the counter terms are obtained. Here the phrase $neutr$ denotes the 
neutral Higgs bosons, $H^0$, $h^0$ and  $A^0$ and  $p$ is the momentum 
transfer. Note that the self energy diagrams do not contribute in this 
scheme. 

\item Integrating out the heavy degrees of freedom, namely $t$ quark, 
$W^{\pm}, H^{\pm}, H^0, h^0$, and $A^0$ bosons in the present case 
and obtaining the effective theory. 

\item Performing the QCD corrections through matching the full theory with 
the effective low energy one at the high scale $\mu=m_{W}$ and evaluating 
the Wilson coefficients from $m_{W}$ down to the lower scale 
$\mu\sim O(m_{b})$. 

\item Obtaining the effective Hamiltonian relevant for the process 
$b\rightarrow s \tau^+\tau^-$ which is given by 
\begin{eqnarray}
{\cal{H}}_{eff}=-4 \frac{G_{F}}{\sqrt{2}} V_{tb} V^{*}_{ts}\aga 
\sum_{i}C_{i}(\mu) O_{i}(\mu)+\sum_{i}C_{Q_i}(\mu) Q_{i}(\mu)\adr 
\, \, ,
\label{hamilton}
\end{eqnarray}
where $O_{i}$ are current-current ($i=1,2$), penguin ($i=3,...,6$),
magnetic penguin ($i=7,8$) and semileptonic ($i=9,10$) operators. 
Here, $C_{i}(\mu)$ are Wilson coefficients normalized at the
scale $\mu $ and given in Appendix B. The additional operators $Q_{i}
(i=1,..,10)$ are due to the NHB exchange diagrams and $C_{Q_i}(\mu)$ are
their Wilson coefficients (see Appendices A and B) . 

\end{itemize}

Therefore the QCD corrected amplitude for the inclusive 
$b\rightarrow s \tau^+ \tau^-$ decay in the model III reads as, 
\begin{eqnarray}
{\cal M} & = & \frac{\alpha_{em} G_F}{ \sqrt{2}\, \pi} V_{tb} V_{ts}^*\Bigg{\{} 
C_9^{eff} (\bar s \gamma_\mu P_L b) \, \bar \tau \gamma_\mu \tau +
C_{10} ( \bar s \gamma_\mu P_L b) \, \bar \tau \gamma_\mu \gamma_5 \tau  \nnb 
\\ & -& 2 C_7^{eff} \frac{m_b}{q^2} (\bar s i \sigma_{\mu \nu} q_\nu P_R b) 
\bar \tau \gamma_\mu \tau + C_{Q_{1}}(\bar s  P_R b) \bar \tau  
\tau +C_{Q_{2}} (\bar s P_R b) \bar \tau \gamma_5 \tau \Bigg{\}}~. 
\label{M1}
\end{eqnarray}

The matrix element  for $B\rightarrow K^* l^+ l^-$ decay can be obtained by 
inserting the inclusive level effective Hamiltonian in eq. (\ref{hamilton}) 
between inital, $B$, and final, $K^*$, hadronic states. The necessary matrix 
elements in this calculation are 
$ \la K^* \vel \bar s \gamma_\mu (1\pm \gamma_5) b \ver B \ra$, 
$\la K^* \vel \bar s i \sigma_{\mu \nu} q^\nu (1+\gamma_5) b \ver B \ra$ 
and $ \la K^* \vel \bar s (1\pm \gamma_5) b \ver B \ra$.
They are calculated by using some non-pertubative methods like QCD sum
rules, light-cone QCD sum rules, etc., and  using the parametrization of 
the form factors as in \cite{Colangelo}, the matrix element of the 
$B\rightarrow K^* \tau^+ \tau^-$ decay is obtained as 
\cite{Alievsum}:
\begin{eqnarray}
{\cal M} &=& -\frac{G \alpha_{em}}{2 \sqrt 2 \pi} V_{tb} V_{ts}^*  
\Bigg\{ \bar {\tau} \gamma^\mu
\tau \left[ 2 A \epsilon_{\mu \nu \rho \sigma} \epsilon^{* \nu} 
p_{K^*}^\rho q^\sigma + i
B_{1} \epsilon^*_\mu - i B_{2} ( \epsilon^* q) 
(p_{B}+p_{K^*})_\mu - 
i B_{3} (\epsilon^* q)q_\mu \right] \nonumber \\
&+& \bar {\tau} \gamma^\mu \gamma_5 \tau \left[ 2 C 
\epsilon_{\mu \nu \rho \sigma}\epsilon^{* \nu} p_{K^*}^\rho q^\sigma + 
i D_{1} \epsilon^*_\mu - i D_{2} (\epsilon^* q) (p_{B}+p_{K^*})_\mu - 
i D_{3} (\epsilon^* q) q_\mu \right] \nonumber \\
&+& i\,\bar {\tau} \, \tau \, F (\epsilon^* q) + i\, 
\bar {\tau} \gamma_5 \tau \, G (\epsilon^* q)   \Bigg\}~,
\label{matr2}
\end{eqnarray}
where $\epsilon^{* \mu}$ is the polarization vector of $K^*$ meson, $p_{B}$ 
and $p_{K^*}$ are four momentum vectors of $B$ and $K^*$ mesons, 
$q=p_B-p_{K^*}$. $A$, $C$, $F$ and $G$, $B_{i}$ and $D_{i}$,  
$i=1,2,3$  are functions of Wilson coefficients and form factors of the
relevant process. Their explicit forms are given in Appendix C. 

Now we are ready to calculate the forward-backward asymmetry of lepton pair, 
CP-violating asymmetry and CP violating asymmetry in forward-backward 
asymmetry for the given process. 

The forward-backward asymmetry  $A_{FB}$ of the lepton pair is a measurable 
physical quantity which provides important clues to test the theoretical 
models used. Using the definition of differential  $A_{FB}$  
\begin{eqnarray}
A_{FB}& = & \frac{ \int^{1}_{0}dz \frac{d \Gamma }{dz} - 
\int^{0}_{-1}dz \frac{d \Gamma }{dz}}{\int^{1}_{0}dz 
\frac{d \Gamma }{dz}+ \int^{0}_{-1}dz \frac{d \Gamma }{dz}}
\label{AFB1} 
\end{eqnarray}
with $z=\cos \theta$, where   $\theta$ is the angle between the momentum of 
B-meson and that of $\tau^{-}$ in the center of mass frame of the 
dileptons $\tau^{+}\tau^{-}$, we get
\begin{eqnarray}
A_{FB}=\frac{\int\, ds\,E(s)}{\int\, ds\,D(s)}. 
\label{AFB2}
\end{eqnarray}
Here,
\begin{eqnarray}
E(s) &=& 6 \,m_B\, \lambda\,v^2\, \Bigg \{ 
\frac{1}{m_b\,r\,(r-1)}\,\Bigg (
4\, m_B\,m_{\tau}\, Re(C_7^{eff *}\, C_{Q_1}) 
\Big (
m_b (\sqrt{r}-1)\, A_2\,(q^2) \nonumber \\ &+& 
m_b (\sqrt{r}+1)\, A_1\,(q^2)-2\, m_B\,s\, T_3\,(q^2)
\Big )  \Big ( 
(r-1)\,(3\,r-s+1)\,T_2\,(q^2)+(r^2+(s-1)^2 \nonumber \\
&-& 2\,r\,(s+1))\,T_3\,(q^2)
\Big ) \Bigg ) \nonumber \\
&-& \frac{1}{m_b^2\,r\,(1+\sqrt {r})}\,
\Bigg (
m_B^2\,m_{\tau}\, Re(C_{9}^{eff *}\, C_{Q_1}) 
\Big (
m_b (\sqrt{r}-1)\, A_2\,(q^2)+m_b (\sqrt{r}+1)\, A_1\,(q^2)\nonumber \\ 
&-&2\, m_B\,s\, T_3\,(q^2)
\Big )  \Big ( (1+\sqrt{r})^2\,(r+s-1)\,A_1\,(q^2) + \lambda \,A_2\, (q^2) 
\Big ) \Bigg ) \nonumber \\ &+& 
8\,C_{10}\, \Bigg (
-2\, m_B\, m_b\,(\sqrt{r}-1)\, Re\,(C_7^{eff})\, T_2\,(q^2)\,
V\,(q^2)\nonumber \\ &+&
A_1\,(q^2) \Big (2\, m_b\,m_B (\sqrt{r}+1)\, Re\,(C_7^{eff})\, T_1\,(q^2)+
m_B^2\,s\, Re(C_{9}^{eff})\, V (q^2) \Big ) \Bigg ) \Bigg \}
\label{AFBnum}
\end{eqnarray}
\begin{eqnarray}
D(s) &=& \sqrt{\lambda}\,v\,\Bigg \{
\frac{32}{m_B\,s^2} m_b^2\, |C_{7}|^2\,  
(2\,m^2_{\tau}+m^2_{B_s}\,s ) \Bigg ( \frac{2\, s}{r\, (r-1)} (1+3\,r-s) 
(T_2 (q^2) \,T_3 (q^2)\,\lambda )\nonumber \\ &+& 
8\, T_1^2 (q^2) \, \lambda  
+ \frac{T_3^2 (q^2)\, s \, \lambda^2}{r\,(r-1)^2} + \frac{1}{r}\,T_2^2 (q^2) 
\Big ( 12\,(r-1)^2\,r-(4\,r-s)\,\lambda \Big ) \Bigg ) \nonumber \\
&+&
\frac{2}{(1+\sqrt{r})^2\,r\,s} m_B |C_{9}^{eff}|^2  
(2\,m^2_{\tau}+m^2_{B_s}\,s ) \Bigg ( 2\, A_1 (q^2)\,A_2(q^2)\,
(1+\sqrt{r})^2\,(r+s-1)\,\lambda\nonumber \\ &+& 
A^2_1 (q^2)\,(1+\sqrt{r})^4\,(12\,r\,s+\lambda)+\lambda\,(8\, r\,s\,V^2 (q^2) 
\, + A^2_2 (q^2)\,\lambda) \Bigg ) \nonumber \\ &+&
2\, C^2_{10}\, m_B\Bigg (
\frac{1}{r\,s}\Bigg ( 2\,A_1(q^2)\, A_2(q^2) (2\,m^2_{\tau} (r-2\,s-1)+
m^2_{B_s}\,s\, (r+s-1) )\lambda \Big ) \nonumber \\ &-&
\frac{1}{m_b\,r} \Big ( 24\,m_B\,m^2_{\tau} (A_2 (q^2)\,(-1+\sqrt{r})+
A_1 (q^2)\,(1+\sqrt{r}))\,T_3 (q^2)\,\lambda \Big ) \nonumber \\
&+& \frac{24}{m_b^2\,r} \,m^2_{B_s}\,m^2_{\tau}\,\lambda\,s\,T_3 (q^2)+
\frac{8}{(1+\sqrt{r})^2} (m^2_{B_s}\,s-4\,m^2_{\tau})\, V^2(q^2)\,\lambda) 
\nonumber \\ &+& 
\frac{\lambda}{(1+\sqrt{r})^2\,r\,s} A^2_2 (q^2) \Big
(m^2_{B_s}\,s\,\lambda+ 2\,m^2_{\tau} (6\,s\,(1+r)-3\,s^2+\lambda) ) \Big )
\nonumber \\ &+& \frac{1}{r\,s} \Big ( (1+\sqrt{r})^2\, A^2_1 (q^2) 
(m^2_{B_s}\,s (12\,r\,s+\lambda)+m^2_{\tau} (-48\,r\,s+2\,\lambda) )\Big )
\Bigg ) \nonumber \\ &+& \!\!
\frac{3\,m^3_{B}\,\lambda}{m_b^4\,r}\,|C_{Q_1}|^2\, \Big (m^2_{B_s}\,s-4\,
m^2_{\tau} \Big ) \Big ( A_2 (q^2) \,m_b\, (\sqrt{r}-1)+A_1 (q^2) \,m_b\, 
(\sqrt{r}+1) -2\,T_3 (q^2)\,m_B\,s\,\Big )^2 \nonumber \\ 
&+&
\frac{3\,m_B^5\,\lambda}{m_b^2\,r} \,|C_{Q_2}|^2\,s 
\Big ( A_2 (q^2) \, (\sqrt{r}-1)+A_1 (q^2) \,(\sqrt{r}+1) 
-2\,T_3 (q^2)\,m^2_{B_s}\,\sqrt{r}\,s\,\Big )^2 \nonumber \\ 
&+&
\frac{1}{(1+\sqrt{r})^2\,s}\,Re(C_7^{eff *}\,C_9^{eff})
16\,m_b\,(2\,m^2_{\tau}+m^2_{B_s}\,s) \Bigg (
8 (1+\sqrt{r})\,T_1 (q^2)\,V (q^2)\,\lambda \nonumber \\
&-& \frac{1}{(-1+\sqrt{r})\,r} \Big ( A_2 (q^2)\,( \lambda (r-1)(1+3\,r-s)
\,T_2 (q^2)+\lambda\,T_3 (q^2))+A_1 (q^2)\,(1+\sqrt{r})^2 \, \nonumber \\ 
&(& (r-1)\,
T_2 (q^2)\, (12 (r-1)\,r-\lambda)+  (r+s-1)\, T_3
(q^2)\,\lambda ) \Big ) \Bigg ) \nonumber \\ &-& 
\frac{12}{m_b^2\,r}\,C_{10}\,Re(C_{Q_2})\, m^3_{B_s}\,m_{\tau} 
\Big (m_b\,( (-1+\sqrt{r})\,A_2 (q^2)+(1+\sqrt{r})\,A_1 (q^2))\nonumber \\ 
&-&2\,m_B\,T_3 (q^2) \Big )  \Big ( A_2 (q^2) (1-\sqrt{r})- 
A_1 (q^2)\, (1+\sqrt{r})+2\,m^2_{B_s}\,\sqrt{r}\,s\,T_3 (q^2)\Big )\,
\lambda   \Bigg \}
\label{AFBden}
\end{eqnarray}
where $\lambda = 1+r^2+s^2 -2 r - 2 s - 2 r s$, $r =
\frac{m_{K^*}^2}{m_B^2}$ and $s=\frac{q^2}{m_B^2}$.

The NHB effects bring new contribution to $A_{FB}$ and we will study those
contributions in the Discusssion part.

The complex Yukawa couplings are the possible source of CP 
violation in the model III. 
In our calculations we neglect all the Yukawa couplings, except 
$\bar{\xi}^{U}_{N,tt}$, $\bar{\xi}^{D}_{N,bb}$ and
$\bar{\xi}^{D}_{N,\tau\tau}$ and choose $\bar{\xi}^{D}_{N,bb}$ complex,
$\bar{\xi}^{D}_{N,bb}=|\bar{\xi}^{D}_{N,bb}|\,e^{i\theta}$ 
(see Discussion part).  Therefore  the CP violation comes from the 
Wilson coefficients $C_7^{eff}$, $C_{Q_1}$ and $C_{Q_2}$. 
Using the definition of $A_{CP}$
\begin{eqnarray}
A_{CP}= \frac{\Gamma (B\rightarrow K^* \tau^+ \tau^-)-
\Gamma (\bar{B}\rightarrow \bar{K^*} \tau^+ \tau^-)}
{\Gamma (B\rightarrow K^* \tau^+ \tau^-)+
\Gamma (\bar{B}\rightarrow \bar{K^*} \tau^+ \tau^-)}
\,\, .
\label{cpvio1}
\end{eqnarray}
we get 
\begin{eqnarray}
A_{CP}= \frac{\int \, ds \, \Omega(s)}{\int \, ds \, \Lambda(s)}
\,\, .
\label{cpvio2}
\end{eqnarray}
where
\begin{eqnarray}
\Omega(s) &=& \frac{m_b\, \alpha^2_e\,G_F^2\,\lambda^2_t}
{384\,\pi^5\,s\,(1+\sqrt {r})^2}\,v\,\sqrt{\lambda}\,
Im (C_7^{eff})\, Im(C_9^{eff})\,(2\,m^2_{\tau}+m^2_{B_s}\,s) \Bigg \{
8\,(\sqrt{r}+1)\,\lambda\,T_1 (q^2)\,V (q^2) \nonumber \\ &-&
\frac{1}{r\,(\sqrt{r}-1)} \Bigg (
\lambda \, A_2 (q^2)\, \Big( (r-1)(3\,r-s+1)\,T_2 (q^2)+\lambda \, 
T_3 (q^2) \Big ) \nonumber \\ &+&  (\sqrt{r}+1)^2\, A_1 (q^2)\, \Big( 
(r-1)\, T_2 (q^2) (12\,(r-1)\,r-\lambda)+ (r+s-1)\,\lambda \,T_3 (q^2) 
\Big ) \Bigg ) \Bigg \}
\label{ACPnum}
\end{eqnarray}
and 
\begin{eqnarray}
\Lambda(s) = D(s)+D_{CP}(s)\,\, .
\label{ACPden}
\end{eqnarray}
Here $D_{CP}(s)$ is the CP conjugate of $D(s)$ which is defined as 
\begin{eqnarray}
D_{CP}(s)= D(s) (\bar{\xi}^D_{N,bb}\rightarrow \bar{\xi}^{D *}_{N,bb}) \,\, .
\end{eqnarray}

The CP violating asymmetry in $A_{FB}$ is also a measurable physical 
quantity and it can give strong clues for the physics beyond the SM. 
This quantity is defined as 
\begin{eqnarray}
A_{CP} (A_{FB})= \frac{A_{FB}-\bar{A}_{FB}}{A_{FB}+\bar{A}_{FB}}
\,\, .
\label{ACPAFB}
\end{eqnarray}
where $\bar{A}_{FB}$ is the CP conjugate of $A_{FB}$ and it is given as 
\begin{eqnarray}
\bar{A}_{FB}= A_{FB} (\bar{\xi}^D_{N,bb}\rightarrow \bar{\xi}^{D *}_{N,bb}) 
\,\, .
\label{AFBCP}
\end{eqnarray}

Note that during the calculations of $A_{CP}$, $A_{FB}$ and $A_{CP}(A_{FB})$ 
we take into account only the second resonance for the LD effects coming from 
the reaction $b \rightarrow s \psi_i \rightarrow s \tau^{+}\tau^{-}$, where 
$i=1,..,6$ and divide the integration region for $s$ into two parts : 
$\frac{4 m^2_{\tau}}{m^2_B}\leq s \leq \frac{(m_{\psi_2}-0.02)^2}{m^2_B}$ 
and $\frac{(m_{\psi_2}+0.02)^2}{m^2_B}\leq s \leq 1$, where
$m_{\psi_2}=3.686\,GeV$ is the mass of the second resonance 
(see Appendix B for LD contributions). 
\section{Discussion}
In the general 2HDM model, the number of free parameters, namely the masses 
of charged and neutral Higgs bosons and complex Yukawa couplings 
($\xi_{ij}^{U,D}$), increases compared to the ones in the SM and model I (II) 
version of 2HDM. The arbitrariness of the numerical values of these
parameters can be removed by using the restrictions coming from the 
experimental measurements. 

Since the neutral Higgs bosons, $h^0$ and $A^0$, can give a large 
contribution to the coefficient $C_7^{eff}$ (see the Appendix of \cite{Alil2} for 
details) which is in contradiction with the CLEO data  \cite{Cleo2}, 
\begin{eqnarray}
Br (B\rightarrow X_s\gamma)= (3.15\pm 0.35\pm 0.32)\, 10^{-4} \,\, ,
\label{br2}
\end{eqnarray}
we take $\bar{\xi}^{D}_{N, ib} \sim 0$ and $\bar{\xi}^{D}_{N, ij}\sim 0$, 
where the indices $i,j$ denote d and s quarks . 
Further we use the constraints \cite{Alil1}, coming from the $\Delta F=2$ 
mixing (here $F=K,B_d,D$) decays, the $\rho$ parameter \cite{Atwood}, and 
the measurement by CLEO Collaboration eq. (\ref{br2}), we get the condition  
for $\bar{\xi}_{N tc}$, $\bar{\xi}_{N tc} << \bar{\xi}^{U}_{N tt}$ and take 
into account only the Yukawa couplings of quarks $\bar{\xi}^{U}_{N,tt}$ and 
$\bar{\xi}^{D}_{N,bb}$. We keep the Yukawa coupling 
$\bar{\xi}^{D}_{N,\tau\tau}$ free and increase this parameter to enhance the 
effects of neutral Higgs bosons. 

In this section, we study the CP parameter $sin\theta$,
the Yukawa coupling $\bar{\xi}^{D}_{N,\tau\tau}$ and the mass ratio
$\frac{m_{h^0}}{m_{A^0}}$ dependencies of the $A_{FB}$, $A_{CP}$ and 
$A_{CP} (A_{FB})$ of the exclusive decay $B\rightarrow K^* \tau^+  \tau^-$, 
restricting $|C_7^{eff}|$ in the region $0.257 \leq |C_7^{eff}| \leq 0.439$
due to the CLEO measurement, eq.(\ref{br2}) (see \cite{Alil1} for details). 
Our numerical calculations based on this restriction and throughout these 
calculations, we use the redefinition
\begin{eqnarray}
\xi^{U,D}=\sqrt{\frac{4 G_F}{\sqrt{2}}} \bar{\xi}^{U,D} \nonumber \,\, , 
\label{xineutr}
\end{eqnarray}
we take $|\frac{\bar{\xi}_{N,tt}^{U}}{\bar{\xi}_{N,bb}^{D}}| <1$, the scale 
$\mu=m_b$, include the LD effects and use the input values 
given in Table (\ref{input}). 
\begin{table}[h]
        \begin{center}
        \begin{tabular}{|l|l|}
        \hline
        \multicolumn{1}{|c|}{Parameter} & 
                \multicolumn{1}{|c|}{Value}     \\
        \hline \hline
        $m_{\tau}$                   & $1.78$ (GeV) \\
        $m_c$                   & $1.4$ (GeV) \\
        $m_b$                   & $4.8$ (GeV) \\
        $\bar{\xi}_{N,bb}^{D}$  &$40\, m_b$  \\
        $\alpha_{em}^{-1}$      & 129           \\
        $\lambda_t$            & 0.04 \\
        $m_{t}$             & $175$ (GeV) \\
        $m_{W}$             & $80.26$ (GeV) \\
        $m_{Z}$             & $91.19$ (GeV) \\
        $m_{H^0}$             & $150$ (GeV) \\
        $m_{h^0}$             & $70$ (GeV) \\  
        $m_{H^{\pm}}$         & $400$ (GeV) \\
        $\Lambda_{QCD}$             & $0.225$ (GeV) \\
        $\alpha_{s}(m_Z)$             & $0.117$  \\
        $sin\theta_W$             & $0.2325$  \\
        \hline
        \end{tabular}
        \end{center}
\caption{The values of the input parameters used in the numerical
          calculations.}
\label{input}
\end{table}

\begin{figure}[htb]
\vskip -3.0truein
\centering
\epsfxsize=6.8in
\leavevmode\epsffile{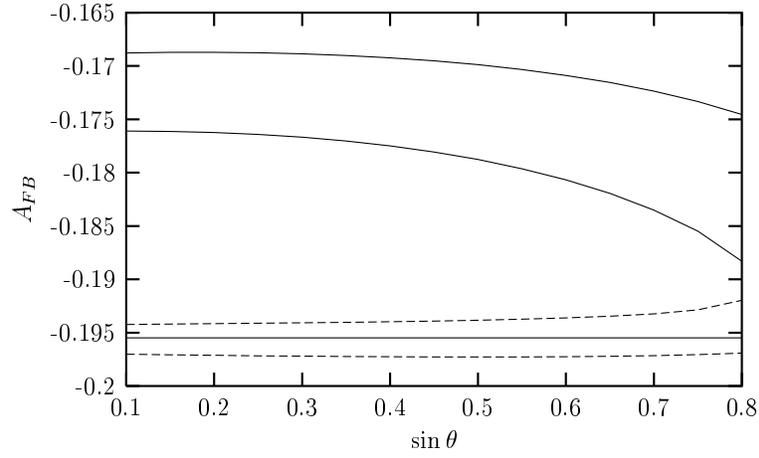}
\vskip -3.0truein
\caption[]{$A_{FB}$ as a function of  $sin\theta$ for $m_{A^0}=80\,GeV$
without NHB effects. Here $A_{FB}$ is restricted in the region between 
solid (dashed) lines for $C_7^{eff} > 0$ ($C_7^{eff} < 0$). Straight line 
corresponds to the SM contribution.} 
\label{AFBsin}
\end{figure}

\begin{figure}[htb]
\vskip -3.0truein
\centering
\epsfxsize=6.8in
\leavevmode\epsffile{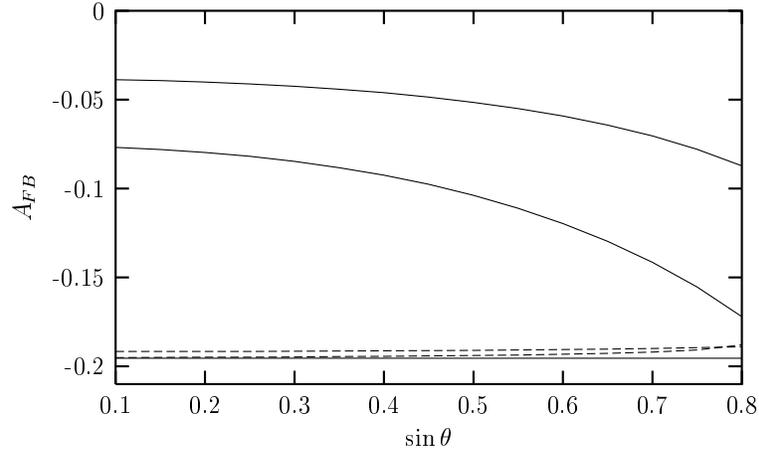}
\vskip -3.0truein
\caption[]{The same as Fig.\ref{AFBsin}, but for $\bar{\xi}_{N,\tau\tau}^{D}=
10\, m_{\tau}$ and including NHB effects.}
\label{AFBNHBsin}
\end{figure}

In Fig. \ref{AFBsin} we present $sin\theta$ dependence of $A_{FB}$ without
NHB effects, for 
$m_{A^0}=80\, GeV$.  Here $A_{FB}$ lies in the region bounded by 
solid (dashed) lines for $C_7^{eff} > 0$ ($C_7^{eff} < 0$). The solid 
straight line shows the SM contribution. In the model III without NHB 
effects, $|A_{FB}|$ is smaller compared to the one in the SM ($0.195$), 
for $C_7^{eff} > 0$, however it is possible to enhance it at the 
order of the magnitude $2\%$ with increasing $sin\theta$. For 
$C_7^{eff} < 0$, $A_{FB}$ is not sensitive to $sin\theta$ and the 
restriction region is narrow. For this case $|A_{FB}|$ can have slightly 
greater values compared to the SM one. Addition of NHB effects (see Fig. 
\ref{AFBNHBsin}) reduces $|A_{FB}|$ for $C_7^{eff} > 0$ almost $30\%$ 
compared to the one without NHB effects. For $C_7^{eff} < 0$, the 
restriction region becomes narrow and  $A_{FB}$ reaches the SM prediction 
for small $sin\theta$.

\begin{figure}[htb]
\vskip -3.0truein
\centering
\epsfxsize=6.8in
\leavevmode\epsffile{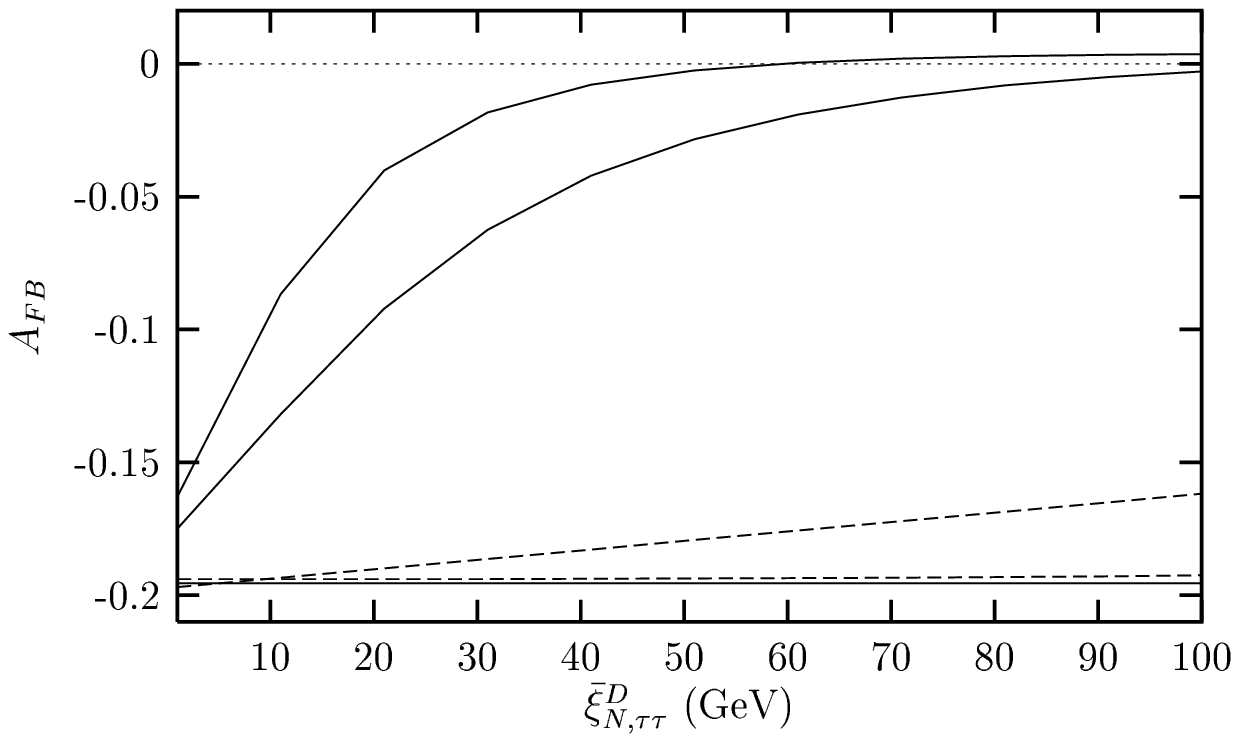}
\vskip -3.0truein
\caption[]{$A_{FB}$ as a function of  $\bar{\xi}^D_{N,\tau\tau}$ for
$sin\theta =0.5$ and $m_{A^0}=80\,GeV$. Here $A_{FB}$ is restricted in the 
region between solid (dashed) lines for $C_7^{eff} > 0$ ($C_7^{eff} < 0$). 
Straight line corresponds to the SM contribution.}
\label{AFBNHBktt}
\end{figure}

Fig. \ref{AFBNHBktt} represent $\bar{\xi}_{N,\tau\tau}^{D}$ dependence of 
$A_{FB}$ for $sin\theta=0.5$ and $m_{A^0}=80\, GeV$.  $|A_{FB}|$ vanishes 
with increasing $\bar{\xi}_{N,\tau\tau}^{D}$ for $C_7^{eff} > 0$. For 
$C_7^{eff} < 0$, $|A_{FB}|$ does not vanish in the given region of 
$\bar{\xi}_{N,\tau\tau}^{D}$  and it stands less than the SM result.  

\begin{figure}[htb]
\vskip -3.0truein
\centering
\epsfxsize=6.8in
\leavevmode\epsffile{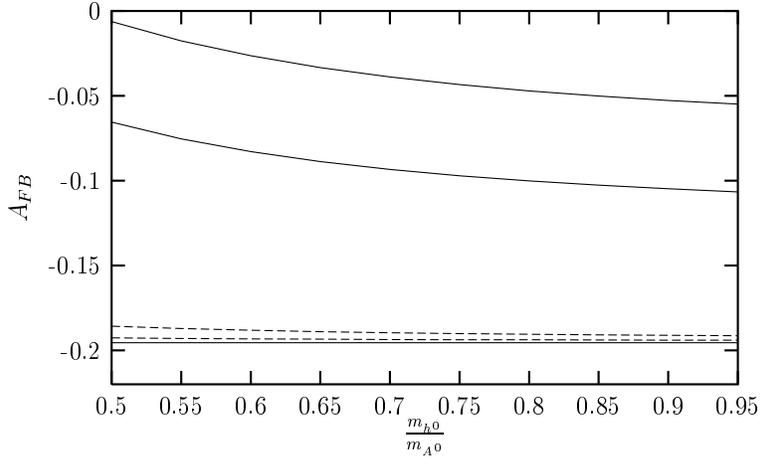}
\vskip -3.0truein
\caption[]{$A_{FB}$ as a function of  $\frac{m_{h^0}}{m_{A^0}}$  for 
$sin\theta =0.5$ and $\bar{\xi}^D_{N,\tau\tau}=10\, m_{\tau}$. Here $A_{FB}$ 
is restricted in the region between solid (dashed) lines for $C_7^{eff} > 0$ 
($C_7^{eff} < 0$). Straight line corresponds to the SM contribution.}
\label{AFBNHBzhA}
\end{figure}

Fig. \ref{AFBNHBzhA} is devoted to the ratio $\frac{m_{h^0}}{m_{A^0}}$ 
dependence of $A_{FB}$ for $sin\theta=0.5$ and $\bar{\xi}_{N,\tau\tau}^{D}
=10\, m_{\tau}$. Increasing values of the ratio causes to increase 
$|A_{FB}|$ for both $C_7^{eff} > 0$ and $C_7^{eff} < 0$. If the masses of
$h^0$ and $A^0$ are far from the degeneracy,  $|A_{FB}|$ becomes small
especially for $C_7^{eff} > 0$.

\begin{figure}[htb]
\vskip -3.0truein
\centering
\epsfxsize=6.8in
\leavevmode\epsffile{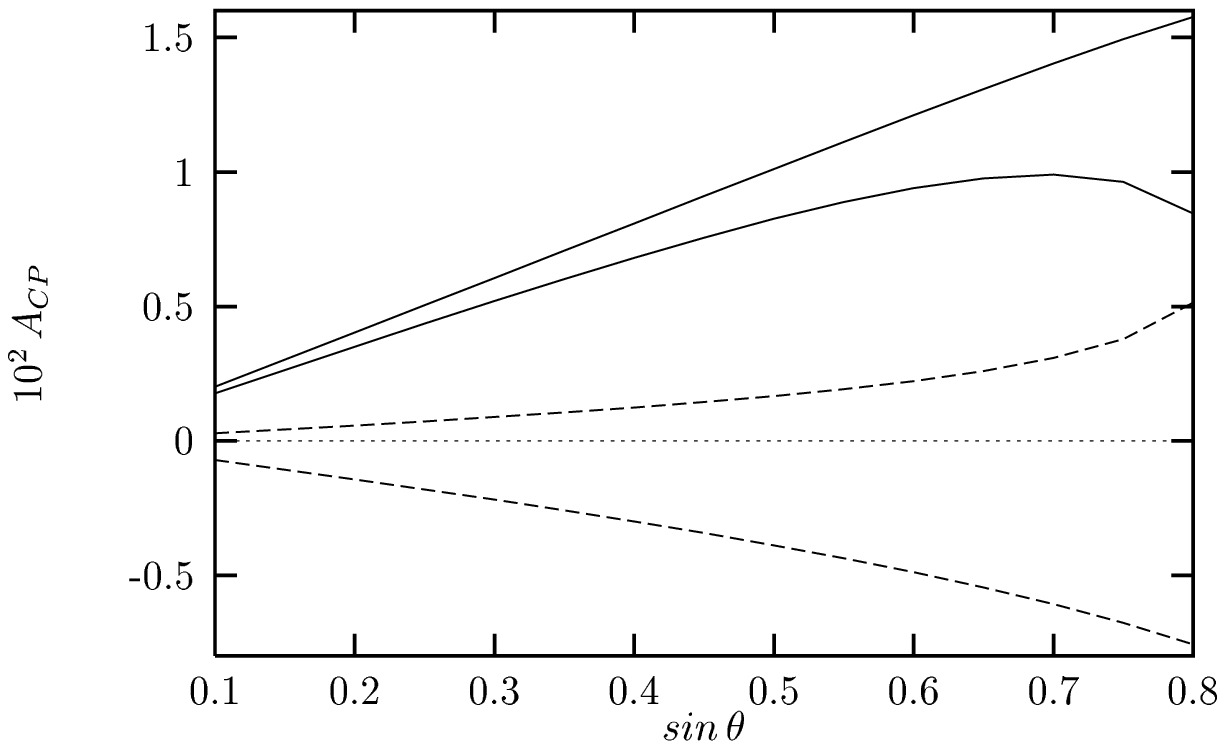}
\vskip -3.0truein
\caption[]{The same as Fig. \ref{AFBsin} but for $A_{CP}$.}
\label{ACPsin}
\end{figure}

\begin{figure}[htb]
\vskip -3.0truein
\centering
\epsfxsize=6.8in
\leavevmode\epsffile{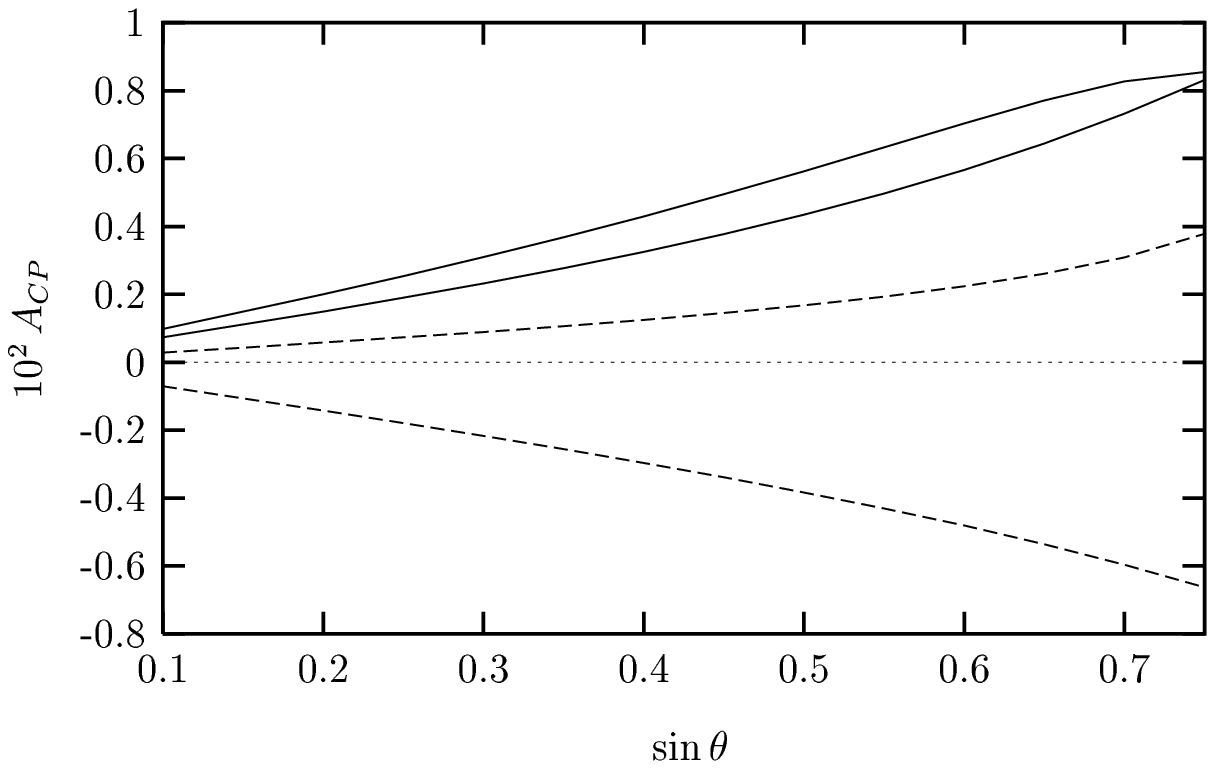}
\vskip -3.0truein
\caption[]{The same as Fig. \ref{AFBNHBsin} but for $A_{CP}$.}
\label{ACPNHBsin}
\end{figure}

\begin{figure}[htb]
\vskip -3.0truein
\centering
\epsfxsize=6.8in
\leavevmode\epsffile{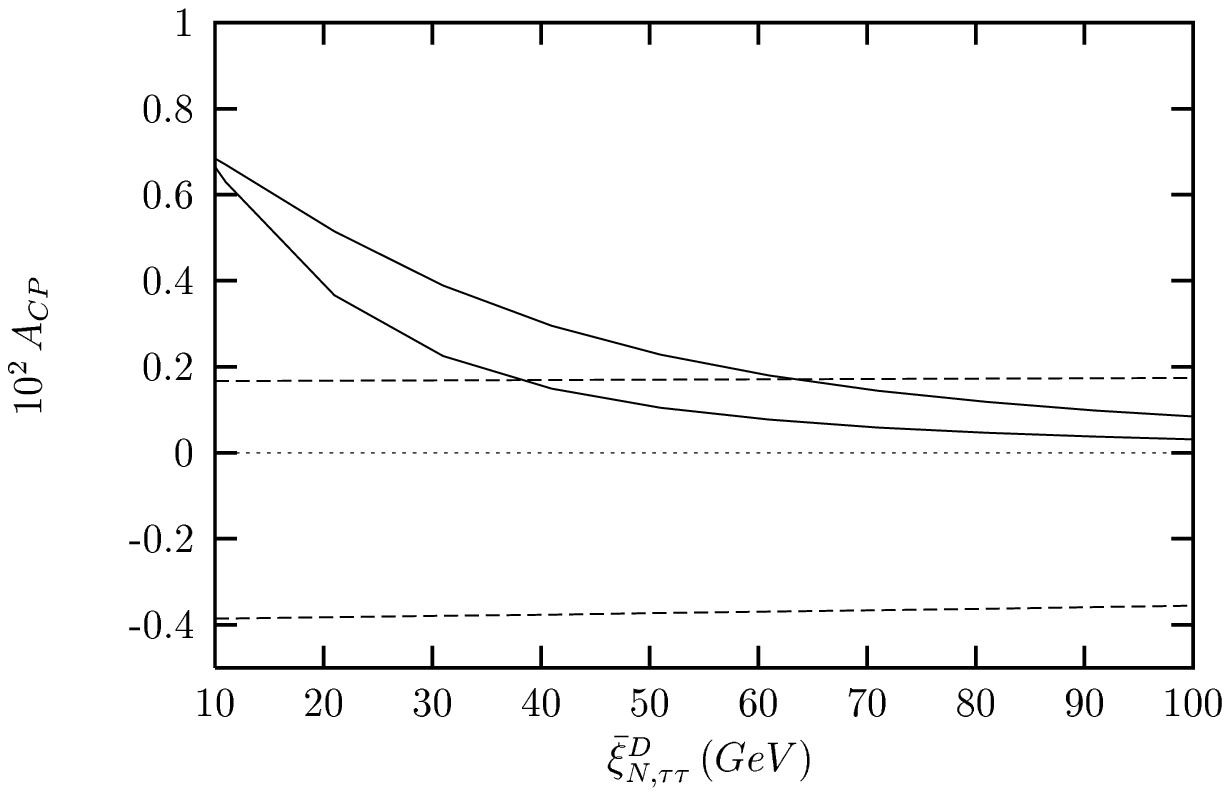}
\vskip -3.0truein
\caption[]{The same as Fig. \ref{AFBNHBktt} but for $A_{CP}$.}
\label{ACPNHBktt}
\end{figure}

Figs \ref{ACPsin}-\ref{ACPNHBktt} represent $A_{CP}$ of the process 
$B \rightarrow K^* \tau^+ \tau^- $. In Fig. \ref{ACPsin} we present 
$sin\theta$ dependence of $A_{CP}$ without NHB effects, for $m_{A^0}=80\, 
GeV$.  Here $A_{CP}$ lies in the region bounded by solid (dashed) lines for 
$C_7^{eff} > 0$ ($C_7^{eff} < 0$). For $C_7^{eff} > 0$, $A_{CP}$ is at the
order of the magnitude of $1\%$ for the intermediate values of $sin\theta$
and its sign does not change in the restriction region.  However 
$A_{CP}$ can have both signs, even vanish for $C_7^{eff} < 0$. With the 
addition of NHB effects (see Fig. \ref{ACPNHBsin}) $A_{CP}$ for $C_7^{eff} 
> 0$ decreases to almost one half of the value we get without NHB effects. 
For $C_7^{eff} < 0$ there is still a decrease in $A_{CP}$.This behavior can 
be seen from the expression eq. (\ref{ACPnum}) since the numerator of the 
$A_{CP}$ ratio is free from NHB effects and their additional contributions 
enter into the expression in the denominator part. Further the restriction 
regions becomes narrow.

Fig. \ref{ACPNHBktt} represent $\bar{\xi}_{N,\tau\tau}^{D}$ dependence of 
$A_{CP}$ for $sin\theta=0.5$ and $m_{A^0}=80\, GeV$.  $A_{CP}$ is sensitive
to the parameter $\bar{\xi}_{N,\tau\tau}^{D}$ and it decreases with 
increasing $\bar{\xi}_{N,\tau\tau}^{D}$ for $C_7^{eff} > 0$. However, for 
$C_7^{eff} < 0$, the dependence of $A_{CP}$  to $\bar{\xi}_{N,\tau\tau}^{D}$ 
is weak.

\begin{figure}[htb]
\vskip -3.0truein
\centering
\epsfxsize=6.8in
\leavevmode\epsffile{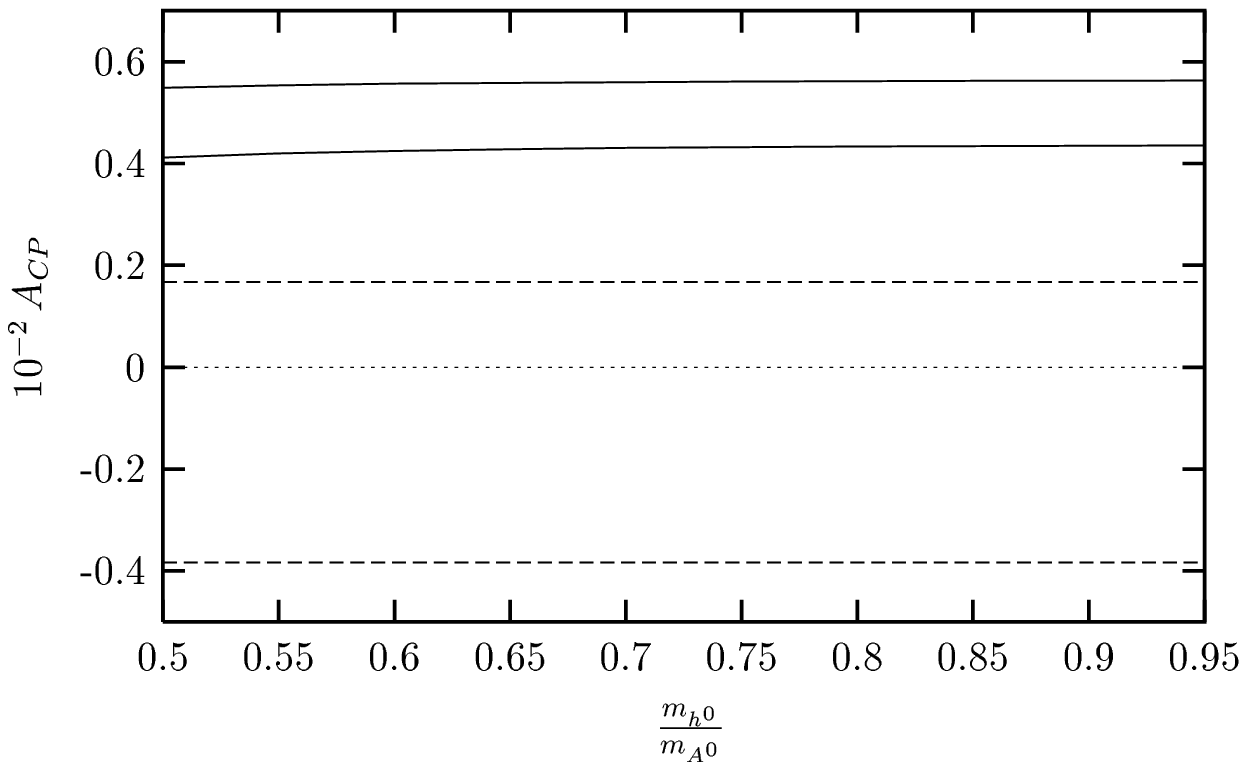}
\vskip -3.0truein
\caption[]{The same as Fig. \ref{AFBNHBzhA} but for $A_{CP}$.}
\label{ACPNHBzhA}
\end{figure}

The ratio $\frac{m_{h^0}}{m_{A^0}}$ dependence of $A_{CP}$ for $sin\theta=
0.5$ and $\bar{\xi}_{N,\tau\tau}^{D}=10\, m_{\tau}$ is presented in Fig. 
\ref{ACPNHBzhA}. As seen from the figure the sensitivity $A_{CP}$ to the
ratio is small, especially for $C_7^{eff} < 0$.

\begin{figure}[htb] 
\vskip -3.0truein   
\centering
\epsfxsize=6.8in    
\leavevmode\epsffile{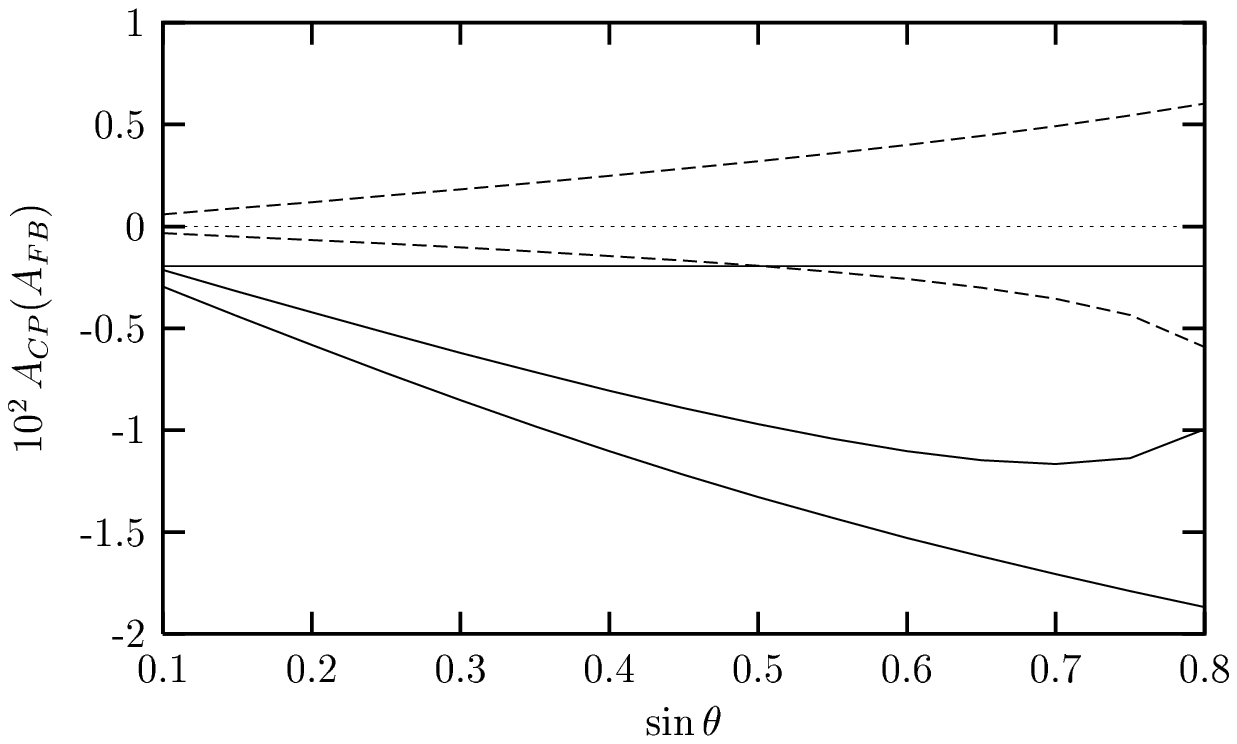}
\vskip -3.0truein   
\caption[]{The same as Fig. \ref{AFBNHBsin} but for $A_{CP}(A_{FB})$.} 
\label{AFBCPNHBsin}    
\end{figure}

\begin{figure}[htb]
\vskip -3.0truein
\centering
\epsfxsize=6.8in
\leavevmode\epsffile{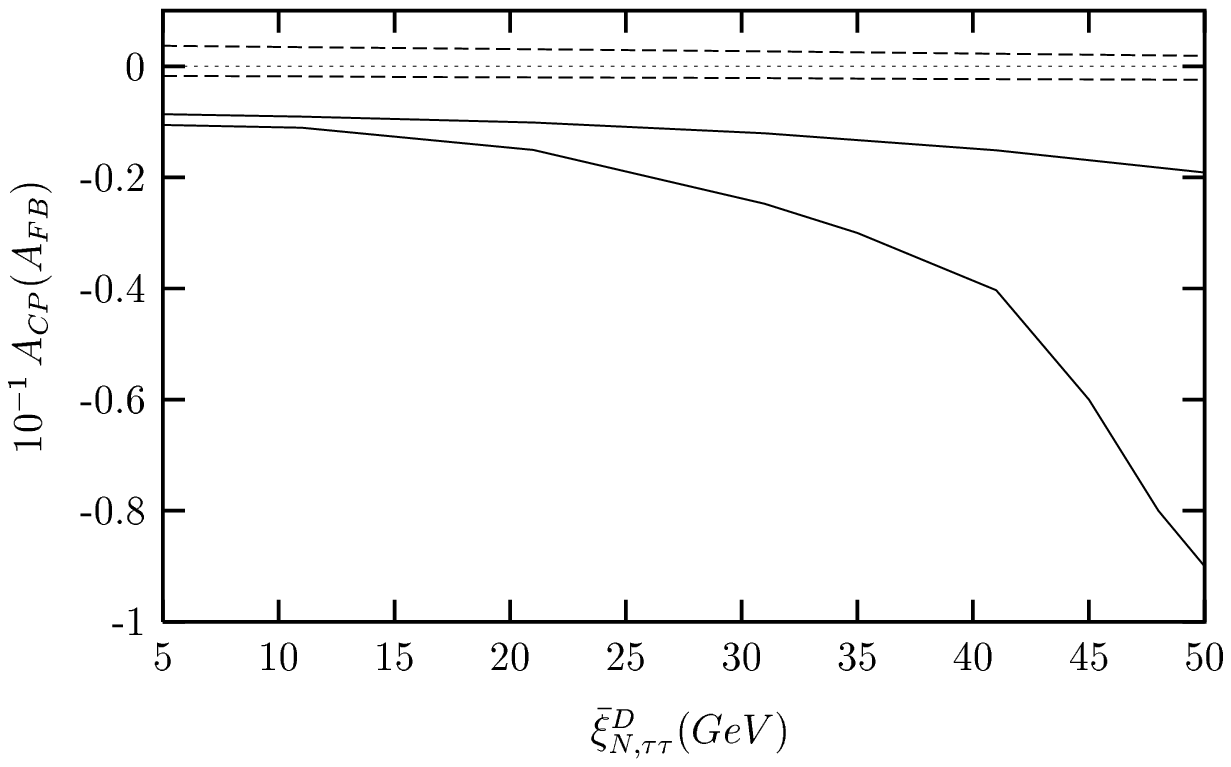}
\vskip -3.0truein
\caption[]{The same as Fig. \ref{AFBNHBktt} but for $A_{CP}(A_{FB})$.}
\label{AFBCPNHBktt} 
\end{figure}

\begin{figure}[htb]
\vskip -3.0truein
\centering
\epsfxsize=6.8in
\leavevmode\epsffile{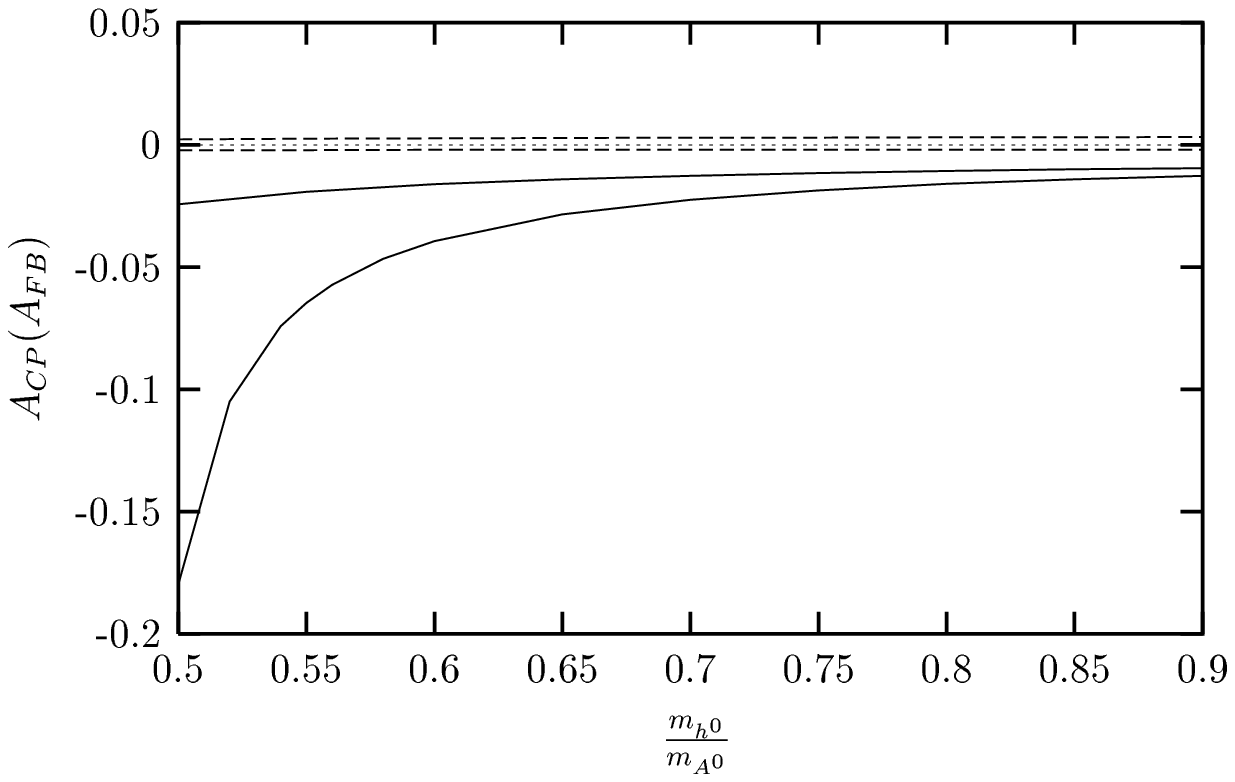}
\vskip -3.0truein
\caption[]{The same as Fig. \ref{AFBNHBzhA} but for $A_{CP}(A_{FB})$.}
\label{AFBCPNHBzhA}
\end{figure}

Finally, we present the CP violating asymmetry in $A_{FB}$ in a series of
figures (Figs. \ref{AFBCPNHBsin}-\ref{AFBCPNHBzhA}). Fig. \ref{AFBCPNHBsin} 
represent $sin\theta$ dependence of $A_{CP}(A_{FB})$ with NHB effects, 
for $m_{A^0}=80\, GeV$ and $\bar{\xi}_{N,\tau\tau}^{D}=10\, m_{\tau}$. 
$A_{CP}(A_{FB})$ is at the order of the magnitude of $1\%$ for the 
intermediate values of $sin\theta$ for $C_7^{eff} > 0$. 
Its sign does not change in the restriction region similar to the $A_{CP}$
of the process under consideration. However $A_{CP}(A_{FB})$ can have both 
signs, even vanish for $C_7^{eff} < 0$. $A_{CP}(A_{FB})$ is sensitive
to the parameter $\bar{\xi}_{N,\tau\tau}^{D}$ especially for the large 
values of $\bar{\xi}_{N,\tau\tau}^{D}$ and $C_7^{eff} > 0$ (see Fig.
\ref{AFBCPNHBktt}). It can reach $10\%$ for $\bar{\xi}_{N,\tau\tau}^{D}=50\,
GeV$. In the case $C_7^{eff} < 0$, $A_{CP}(A_{FB})$ is not sensitive to 
$\bar{\xi}_{N,\tau\tau}^{D}$ and it almost vanishes.

Fig. \ref{AFBCPNHBzhA} is devoted to the ratio $\frac{m_{h^0}}{m_{A^0}}$ 
dependence of $A_{CP}(A_{FB})$ for $sin\theta=0.5$ and 
$\bar{\xi}_{N,\tau\tau}^{D}=10\, m_{\tau}$. Increasing values of the ratio 
causes to increase $|A_{CP}(A_{FB})|$ for $C_7^{eff} > 0$. With the
increasing mass ratio of $h^0$ and $A^0$, $|A_{CP}(A_{FB})|$ can take 
large values. For $C_7^{eff} < 0$, $A_{CP}(A_{FB})$ is not sensitive to 
the mass ratio. 

Now, we would like to summarize our results. 
\begin{itemize}
\item $|A_{FB}|$ for the process under consideration is at the order of 
$10^{-2}$ and smaller compared to the SM one, for $C_{7}^{eff}>0$. It can 
exceed the SM value ($0.195$) for $C_{7}^{eff}<0$. Addition of NHB effects 
decreases its magnitude by $30\%$ (slightly) for $C_{7}^{eff}>0$ 
($C_{7}^{eff}<0$). $A_{FB}$ is sensitive to the parameters $sin\theta$, 
$\bar{\xi}_{N,\tau\tau}^{D}$ and $\frac{m_{h^0}}{m_{A^0}}$ especially for 
$C_{7}^{eff}>0$. Its magnitude decreases (increases) with increasing values 
of $\bar{\xi}_{N,\tau\tau}^{D}$ ($\frac{m_{h^0}}{m_{A^0}}$).

\item $|A_{CP}|$ is at the order of $10^{-2}$. Addition of NHB effects 
decreases its magnitude by $50\%$ (slightly) for $C_{7}^{eff}>0$ 
($C_{7}^{eff}<0$). It has the same sign in the restriction region 
$C_{7}^{eff}>0$ and it can take both signs for $C_{7}^{eff}<0$. $A_{CP}$ 
is sensitive to the parameters $sin\theta$, $\bar{\xi}_{N,\tau\tau}^{D}$  
especially for $C_{7}^{eff}>0$. It decreases with increasing values of 
$\bar{\xi}_{N,\tau\tau}^{D}$. The sensitivity of $A_{CP}$ to the ratio 
$\frac{m_{h^0}}{m_{A^0}}$ is weak.

\item $A_{CP}(A_{FB})$ is at the order of the magnitude of $1\%$ for the 
intermediate values of $sin\theta$ for $C_7^{eff} > 0$. It has the same 
sign in the restriction region $C_{7}^{eff}>0$ and it can take both signs 
for $C_{7}^{eff}<0$. $A_{CP}(A_{FB})$ is sensitive to the parameters 
$\bar{\xi}_{N,\tau\tau}^{D}$ and $\frac{m_{h^0}}{m_{A^0}}$ for 
$C_{7}^{eff}>0$. It increases with increasing values of $\bar{\xi}_
{N,\tau\tau}^{D}$, even reach to $10\%$. Further the increasing values of the 
ratio $\frac{m_{h^0}}{m_{A^0}}$ causes to increase $|A_{CP}(A_{FB})|$.

\end{itemize}

Therefore, the experimental investigation of $A_{FB}$ and $A_{CP}$ and 
$A_{CP}(A_{FB})$ ensure a crucial test for new physics effects beyond the 
SM   and also the sign of $C_{7}^{eff}$.

\newpage
%{\bf \LARGE {Appendix}} \\
\begin{appendix}
\section{The operator basis}
The operator basis in the  2HDM (model III ) for our process  
is \cite{Dai,Grinstein2,Misiak}
\begin{eqnarray}
 O_1 &=& (\bar{s}_{L \alpha} \gamma_\mu c_{L \beta})
               (\bar{c}_{L \beta} \gamma^\mu b_{L \alpha}), \nonumber   \\
 O_2 &=& (\bar{s}_{L \alpha} \gamma_\mu c_{L \alpha})
               (\bar{c}_{L \beta} \gamma^\mu b_{L \beta}),  \nonumber   \\
 O_3 &=& (\bar{s}_{L \alpha} \gamma_\mu b_{L \alpha})
               \sum_{q=u,d,s,c,b}
               (\bar{q}_{L \beta} \gamma^\mu q_{L \beta}),  \nonumber   \\
 O_4 &=& (\bar{s}_{L \alpha} \gamma_\mu b_{L \beta})
                \sum_{q=u,d,s,c,b}
               (\bar{q}_{L \beta} \gamma^\mu q_{L \alpha}),   \nonumber  \\
 O_5 &=& (\bar{s}_{L \alpha} \gamma_\mu b_{L \alpha})
               \sum_{q=u,d,s,c,b}
               (\bar{q}_{R \beta} \gamma^\mu q_{R \beta}),   \nonumber  \\
 O_6 &=& (\bar{s}_{L \alpha} \gamma_\mu b_{L \beta})
                \sum_{q=u,d,s,c,b}
               (\bar{q}_{R \beta} \gamma^\mu q_{R \alpha}),  \nonumber   \\  
 O_7 &=& \frac{e}{16 \pi^2}
          \bar{s}_{\alpha} \sigma_{\mu \nu} (m_b R + m_s L) b_{\alpha}
                {\cal{F}}^{\mu \nu},                             \nonumber  \\
 O_8 &=& \frac{g}{16 \pi^2}
    \bar{s}_{\alpha} T_{\alpha \beta}^a \sigma_{\mu \nu} (m_b R +
m_s L)  
          b_{\beta} {\cal{G}}^{a \mu \nu} \nonumber \,\, , \\  
 O_9 &=& \frac{e}{16 \pi^2}
          (\bar{s}_{L \alpha} \gamma_\mu b_{L \alpha})
              (\bar{\tau} \gamma^\mu \tau)  \,\, ,    \nonumber    \\
 O_{10} &=& \frac{e}{16 \pi^2}
          (\bar{s}_{L \alpha} \gamma_\mu b_{L \alpha})
              (\bar{\tau} \gamma^\mu \gamma_{5} \tau)  \,\, ,    \nonumber  \\
Q_1&=&   \frac{e^2}{16 \pi^2}(\bar{s}^{\alpha}_{L}\,b^{\alpha}_{R})\,(\bar{\tau}\tau ) 
\nnb  \\ 
Q_2&=&    \frac{e^2}{16 \pi^2}(\bar{s}^{\alpha}_{L}\,b^{\alpha}_{R})\,
(\bar{\tau} \gamma_5 \tau ) \nnb \\
Q_3&=&    \frac{g^2}{16 \pi^2}(\bar{s}^{\alpha}_{L}\,b^{\alpha}_{R})\,
\sum_{q=u,d,s,c,b }(\bar{q}^{\beta}_{L} \, q^{\beta}_{R} ) \nnb \\
Q_4&=&  \frac{g^2}{16 \pi^2}(\bar{s}^{\alpha}_{L}\,b^{\alpha}_{R})\,
\sum_{q=u,d,s,c,b } (\bar{q}^{\beta}_{R} \, q^{\beta}_{L} ) \nnb \\
Q_5&=&   \frac{g^2}{16 \pi^2}(\bar{s}^{\alpha}_{L}\,b^{\beta}_{R})\,
\sum_{q=u,d,s,c,b } (\bar{q}^{\beta}_{L} \, q^{\alpha}_{R} ) \nnb \\
Q_6&=&   \frac{g^2}{16 \pi^2}(\bar{s}^{\alpha}_{L}\,b^{\beta}_{R})\,
\sum_{q=u,d,s,c,b } (\bar{q}^{\beta}_{R} \, q^{\alpha}_{L} ) \nnb \\
Q_7&=&   \frac{g^2}{16 \pi^2}(\bar{s}^{\alpha}_{L}\,\sigma^{\mu \nu} \, 
b^{\alpha}_{R})\,
\sum_{q=u,d,s,c,b } (\bar{q}^{\beta}_{L} \, \sigma_{\mu \nu } 
q^{\beta}_{R} ) \nnb \\
Q_8&=&    \frac{g^2}{16 \pi^2}(\bar{s}^{\alpha}_{L}\,\sigma^{\mu \nu} 
\, b^{\alpha}_{R})\,
\sum_{q=u,d,s,c,b } (\bar{q}^{\beta}_{R} \, \sigma_{\mu \nu } 
q^{\beta}_{L} ) \nnb \\ 
Q_9&=&   \frac{g^2}{16 \pi^2}(\bar{s}^{\alpha}_{L}\,\sigma^{\mu \nu} 
\, b^{\beta}_{R})\,
\sum_{q=u,d,s,c,b }(\bar{q}^{\beta}_{L} \, \sigma_{\mu \nu } 
q^{\alpha}_{R} ) \nnb \\
Q_{10}&= & \frac{g^2}{16 \pi^2}(\bar{s}^{\alpha}_{L}\,\sigma^{\mu \nu} \, 
b^{\beta}_{R})\,
\sum_{q=u,d,s,c,b }(\bar{q}^{\beta}_{R} \, \sigma_{\mu \nu } q^{\alpha}_{L} )
\label{op1}
\end{eqnarray}
where $\alpha$ and $\beta$ are $SU(3)$ colour indices and 
${\cal{F}}^{\mu \nu}$ and ${\cal{G}}^{\mu \nu}$ are the field strength 
tensors of the electromagnetic and strong interactions, respectively. Note 
that there are also flipped chirality partners of these operators, which 
can be obtained by interchanging $L$ and $R$ in the basis given above in 
model III. However, we do not present them here since corresponding  Wilson 
coefficients are negligible.
\section{The Initial values of the Wilson coefficients.}
The initial values of the Wilson coefficients for the relevant process 
in the SM are \cite{Grinstein2}
\begin{eqnarray}
C^{SM}_{1,3,\dots 6}(m_W)&=&0 \nonumber \, \, , \\
C^{SM}_2(m_W)&=&1 \nonumber \, \, , \\
C_7^{SM}(m_W)&=&\frac{3 x_t^3-2 x_t^2}{4(x_t-1)^4} \ln x_t+
\frac{-8 x_t^3-5 x_t^2+7 x_t}{24 (x_t-1)^3} \nonumber \, \, , \\
C_8^{SM}(m_W)&=&-\frac{3 x_t^2}{4(x_t-1)^4} \ln x_t+
\frac{-x_t^3+5 x_t^2+2 x_t}{8 (x_t-1)^3}\nonumber \, \, , \\ 
C_9^{SM}(m_W)&=&-\frac{1}{sin^2\theta_{W}} B(x_t) +
\frac{1-4 \sin^2 \theta_W}{\sin^2 \theta_W} C(x_t)-D(x_t)+
\frac{4}{9}, \nonumber \, \, , \\
C_{10}^{SM}(m_W)&=&\frac{1}{\sin^2\theta_W}
(B(x_t)-C(x_t))\nonumber \,\, , \\
C_{Q_i}^{SM}(m_W) & = & 0~~~ i=1,..,10~.
\end{eqnarray}
and for the additional part due to charged Higgs bosons are 
\begin{eqnarray}
C^{H}_{1,\dots 6 }(m_W)&=&0 \nonumber \, , \\
C_7^{H}(m_W)&=& Y^2 \, F_{1}(y_t)\, + \, X Y \,  F_{2}(y_t) 
\nonumber  \, \, , \\
C_8^{H}(m_W)&=& Y^2 \,  G_{1}(y_t) \, + \, X Y \, G_{2}(y_t) 
\nonumber\, \, , \\
C_9^{H}(m_W)&=&  Y^2 \,  H_{1}(y_t) \nonumber  \, \, , \\
C_{10}^{H}(m_W)&=& Y^2 \,  L_{1}(y_t)  
\label{CH} \, \, , 
\end{eqnarray}
where 
\bea
X & = & \frac{1}{m_{b}}~~~\left(\bar{\xi}^{D}_{N,bb}+\bar{\xi}^{D}_{N,sb}
\frac{V_{ts}}{V_{tb}} \right) ~~,~~ \nnb \\
Y & = & \frac{1}{m_{t}}~~~\left(\bar{\xi}^{U}_{N,tt}+\bar{\xi}^{U}_{N,tc}
\frac{V^{*}_{cs}}{V^{*}_{ts}} \right) ~~,~~
\eea
The NHB effects bring new operators and the corresponding Wilson
coefficients read as 
\bea
\!\!\!\!\!\!\!\!\!\!\!\!\!\!\!\!\!\!\!\!\!\!\!\!\!\!\!\!\!\!\!\!\!\!\!\!\!\!\!
\!\!\!\!\!\!\!\!\!\!\!\!\!\!\!\!\!\!\!\!\!\!\!\!\!\!\!\!\!\!\!\!\!\!\!\!\!\!\!
C^{A^{0}}_{Q_{2}}((\bar{\xi}^{U}_{N,tt})^{3}) =
\frac{\bar{\xi}^{D}_{N,\tau \tau}(\bar{\xi}^{U}_{N,tt})^{3}
m_{b} y_t (\Theta_5 (y_t)z_A-\Theta_1 (z_{A},y_t))}{32 \pi^{2}m_{A^{0}}^{2} 
m_{t} \Theta_1 (z_{A},y_t) \Theta_5 (y_t)} , \nnb
\eea
\bea
C^{A^{0}}_{Q_{2}}((\bar{\xi}^{U}_{N,tt})^{2})=\frac{\bar{\xi}^{D}_{N,\tau\tau}
(\bar{\xi}^{U}_{N,tt})^{2}
\bar{\xi}^{D}_{N,bb}}{32 \pi^{2}  m_{A^{0}}^{2}}\Big{(} 
\frac{(y_t (\Theta_1 (z_{A},y_t) - \Theta_5 (y_t) (xy+z_A))-
2 \Theta_1 (z_{A},y_t) \Theta_5 (y_t)   \ln [\frac{z_A \Theta_5 (y_t)}
{ \Theta_1 (z_{A},y_t)}]}{ \Theta_1 (z_{A},y_t) \Theta_5 (y_t)}\Big{)}, \nnb
\eea
\bea
C^{A^{0}}_{Q_{2}}(\bar{\xi}^{U}_{N,tt}) &=& 
\frac{g^2\bar{\xi}^{D}_{N,\tau\tau}\bar{\xi}^{U}_{N,tt} m_b
x_t}{64 \pi^2 m_{A^{0}}^{2}  m_t } \Bigg{(}\frac{2}{\Theta_5 (x_t)}
- \frac{xy x_t+2 z_A}{\Theta_1 (z_{A},x_t)}-2
\ln [\frac{z_A \Theta_5(x_t)}{ \Theta_1 (z_{A},x_t)}]\nnb \\ &
&\!\!\!\!\!\!\!\!\!\!\!\!\!\!\!\!\!\!\!\!\!\!\!- x y x_t y_t(\frac{(x-1) x_t 
(y_t/z_A-1)-(1+x)y_t)}{(\Theta_6 -(x-y)(x_t -y_t))(\Theta_3
(z_A)+(x-y)(x_t-y_t)z_A)}-
\frac{x (y_t+x_t(1-y_t/z_A))-2 y_t }{\Theta_6 \Theta_3 (z_A)}) \Bigg{)} \nnb
\eea
\bea
\!\!\!C^{A^{0}}_{Q_{2}}(\bar{\xi}^{D}_{N,bb}) =
\frac{g^2\bar{\xi}^{D}_{N,\tau\tau}\bar{\xi}^{D}_{N,bb}}{64 \pi^2 m^2_{A^{0}} }
\Big{(}1-
\frac{x^2_t y_t+2 y (x-1)x_t y_t-z_A (x^2_t+\Theta_6)}{ \Theta_3 (z_A)}+
\frac{x^2_t (1-y_t/z_A)}{\Theta_6}+2 \ln [\frac{z_A \Theta_6}{ \Theta_2 (z_{A},x)}]
\Big{)}\nnb 
\eea
\bea
\lefteqn{\!\!\!\!\!\!\!\!\!\!\!\!\!\!\!\!\!\!\!\!\!C^{H^{0}}_{Q_{1}}
((\bar{\xi}^{U}_{N,tt})^{2}) = 
\frac{g^2 (\bar{\xi}^{U}_{N,tt})^2 m_b m_{\tau}
}{64 \pi^2 m^2_{H^{0}} m^2_t } \Bigg{(}
\frac{x_t (1-2 y) y_t}{\Theta_5 (y_t)}+\frac{(-1+2 \cos^2 \theta_W) (-1+x+y) 
y_t} {\cos^2 \theta_W \Theta_4 (y_t)} } \nnb
\\ & &
+\frac{z_H (\Theta_1 (z_H,y_t) x y_t + 
\cos^2 \theta_W \,(-2 x^2 (-1+x_t) y y^2_t+x x_t y y^2_t-\Theta_8 z_H))}
{\cos^2 \theta_W \Theta_1 (z_H,y_t) \Theta_7 }\Bigg{)} ,             
\label{NHB}
\eea
\bea
\lefteqn{\!\!\!\!\!\!\!\!\!\!\!\!\!\!\!\!\!\!\!\!\!\!\!\!\!\!
\!\!\!\!\!\!\!\!\!\!\!\!\!\!\!\!\!\!\!\!\!C^{H^{0}}_{Q_{1}}
(\bar{\xi}^{U}_{N,tt}) 
=\frac{g^2 \bar{\xi}^{U}_{N,tt} \bar{\xi}^{D}_{N,bb} 
m_{\tau}}{64 \pi^2 m^2_{H^{0}} m_t } \Bigg{(}
\frac{(-1+2 \cos^2 \theta_W)\, y_t}{\cos^2 \theta_W \, \Theta_4 (y_t)}-
\frac{x_t y_t}{\Theta_5 (y_t)}+\frac {x_t y_t(x y-z_H)}
{\Theta_1 (z_H,y_t)} } 
\nnb
\\ & & 
+\frac{(-1+2 \cos^2 \theta_W) 
y_t z_H}{\cos^2\theta_W \Theta_7}-2 x_t\, \ln \Bigg{[}
\frac{\Theta_5 (y_t) z_H} {\Theta_1 (z_H,y_t)} \Bigg{]} \Bigg{)}  ,
\nnb
\eea
\bea
\lefteqn{ C^{H^0}_{Q_{1}}(g^4) =-\frac{g^4 m_b m_{\tau} x_t}
{128 \pi^2 m^2_{H^{0}} m^2_t} 
\Bigg{(} -1+\frac{(-1+2x) x_t}{\Theta_5 (x_t) + y (1-x_t)}+
\frac{2 x_t (-1+ (2+x_t) y)}{\Theta_5 (x_t)} } \nnb \\
& & 
-\frac{4 \cos^2 \theta_W (-1+x+y)+ x_t(x+y)} {\cos^2 \theta_W 
\Theta_4 (x_t)} +\frac{x_t (x (x_t (y-2 z_H)-4 z_H)+2 z_H)} {\Theta_1 
(z_H,x_t)} \nnb \\ 
& &
+\frac{y_t ( (-1+x) x_t z_H+\cos^2 \theta_W ( (3 x-y) z_H+x_t 
(2 y (x-1)- z_H (2-3 x -y))))}{\cos^2 \theta_W (\Theta_3 (z_H)+x 
(x_t-y_t) z_H)} \nnb
\\ & & 
+ 2\, ( x_t \ln \Bigg{[} \frac{\Theta_5 (x_t) z_H}{\Theta_1 (z_H,x_t)} 
\Bigg{]}+ \ln \Bigg{[} \frac{x(y_t-x_t) z_H-\Theta_3 (z_H)} {(\Theta_5 (x_t)+ 
y (1-x_t) y_t z_H} \Bigg{]} )\Bigg{)}  ,\nnb  
\eea
\bea
C^{h_0}_{Q_1}((\bar{\xi}^U_{N,tt})^3) &=&
-\frac{\bar{\xi}^D_{N,\tau\tau} (\bar{\xi}^U_{N,tt})^3 m_b y_t}
{32 \pi^2 m_{h^0}^2 m_t \Theta_1 (z_h,y_t) \Theta_5 (y_t)}
 \Big{(} \Theta_1 (z_h,y_t) (2 y-1) + \Theta_5 (y_t) (2 x-1) z_h \Big{)} \nnb 
\eea
\bea
C^{h_0}_{Q_1}((\bar{\xi}^U_{N,tt})^2)=
\frac{\bar{\xi}^D_{N,\tau\tau} (\bar{\xi}^U_{N,tt})^2 }
{32 \pi^2  m_{h^0}^2  } \Bigg{(}
\frac{ (\Theta_5 (y_t) z_h (y_t-1)(x+y-1)-\Theta_1 (z_h,y_t)( \Theta_5(y_t)+y_t )
}{\Theta_1 (z_h)\Theta_5(y_t)}- 2 \ln \Bigg{[} \frac{z_h \Theta_5
(y_t)}{\Theta_1 (z_h)} \Bigg{]} \Bigg{)}\nnb 
\eea
\bea
C^{h^0}_{Q_{1}}(\bar{\xi}^{U}_{N,tt}) & = & -\frac{g^2
\bar{\xi}^{D}_{N,\tau\tau}\bar{\xi}^{U}_{N,tt} m_b x_t}{64 \pi^2 m^2_{h^{0}} 
m_t} \Bigg{(}\frac{2 (-1+(2+x_t) y)}{\Theta_5 (x_t)}-\frac{x_t
(x-1)(y_t-z_h)}{\Theta'_2 (z_h)}+2 \ln \Bigg{[}\frac{z_h \Theta_5
(x_t)}{\Theta_1 (z_h,x_t)} \Bigg{]} \nnb \\ & + & \frac{x (x_t (y-2 z_h)-
4 z_h)+2 z_h}{\Theta_1 (z_h,x_t)}  -  \frac{(1+x) y_t z_h}{x y x_t y_t+z_h
((x-y)(x_t-y_t)- \Theta_6)} \nnb \\ 
& + & 
\frac{\Theta_9 + y_t z_h ( (x-y)(x_t-y_t)-\Theta_6 )(2
x-1)}{z_h \Theta_6 (\Theta_6 -(x-y)(x_t-y_t))}+\frac{x (y_t z_h + x_t
(z_h-y_t))-2 y_t z_h}{\Theta_2 (z_h)} \Bigg{)}, \nnb
\eea
\bea
C^{h^0}_{Q_{1}}(\bar{\xi}^{D}_{N,bb}) &  = & -\frac{g^2
\bar{\xi}^{D}_{N,\tau\tau}\bar{\xi}^{D}_{N,bb}}{64 \pi^2 m^2_{h^0} }
\Bigg{(}\frac{y x_t y_t (x x^2_t(y_t-z_h)+\Theta_6 z_h
(x-2))}{z_h\Theta_2 (z_h)\Theta_6 }+2 \ln \Bigg{[}\frac{\Theta_6}{x_t y_t} \Bigg{]}
+2 \ln \Bigg{[}\frac{x_t y_t z_h}{\Theta_2 (z_h)} \Bigg{]}
\Bigg{)} \nnb 
\eea
where 
\bea
\Theta_1 (\omega , \lambda ) & = & -(-1+y-y \lambda ) \omega -x (y \lambda
+\omega - \omega \lambda ) \nnb \\
\Theta_2 (\omega ) & = &  (x_t +y (1-x_t)) y_t \omega - x x_t (y
y_t+(y_t-1) \omega)   \nnb \\
\Theta^{\prime}_2  (\omega ) & = & \Theta_2 (\omega , x_t \leftrightarrow y_t)    \nnb \\
\Theta_3 (\omega) & = & (x_t (-1+y)-y ) y_t \omega +
x x_t (y y_t+\omega(-1+y_t)) \nnb \\
\Theta_4 (\omega) & = & 1-x +x  \omega  \nnb \\
\Theta_5 (\lambda) & = & x + \lambda (1-x) \nnb \\
\Theta_6  & = & (x_t +y  (1-x_t))y_t +x x_t  (1-y_t) \nnb \\
\Theta_7  & = & (y (y_t -1)-y_t) z_H+x (y y_t + (y_t-1) z_H ) \\ 
\Theta_8  & = & y_t (2 x^2 (1+x_t) (y_t-1) +x_t (y(1-y_t)+y_t)+x
(2(1-y+y_t) \nnb \\ & + & x_t (1-2 y (1-y_t)-3 y_t))) \nnb \\
\Theta_9  & = & -x^2_t (-1+x+y)(-y_t+x (2 y_t-1)) (y_t-z_h)-x_t y_t z_h
(x(1+2 x)-2 y) \nnb \\ & + & y^2_t (x_t (x^2 -y (1-x))+(1+x) (x-y) z_h) 
\nnb
\eea
and
\begin{eqnarray}
& & x_t=\frac{m_t^2}{m_W^2}~~~,~~~y_t=
\frac{m_t^2}{m_{H^{\pm}}}~~~,~~~z_H=\frac{m_t^2}{m^2_{H^0}}~~~,~~~
z_h=\frac{m_t^2}{m^2_{h^0}}~~~,~~~ z_A=\frac{m_t^2}{m^2_{A^0}}~~~,~~~ \nnb
\end{eqnarray}
The explicit forms of the functions $F_{1(2)}(y_t)$, $G_{1(2)}(y_t)$, 
$H_{1}(y_t)$ and $L_{1}(y_t)$ in eq.(\ref{CH}) are given as
\begin{eqnarray}
F_{1}(y_t)&=& \frac{y_t(7-5 y_t-8 y_t^2)}{72 (y_t-1)^3}+
\frac{y_t^2 (3 y_t-2)}{12(y_t-1)^4} \,\ln y_t \nonumber  \,\, , 
\\ 
F_{2}(y_t)&=& \frac{y_t(5 y_t-3)}{12 (y_t-1)^2}+
\frac{y_t(-3 y_t+2)}{6(y_t-1)^3}\, \ln y_t 
\nonumber  \,\, ,
\\ 
G_{1}(y_t)&=& \frac{y_t(-y_t^2+5 y_t+2)}{24 (y_t-1)^3}+
\frac{-y_t^2} {4(y_t-1)^4} \, \ln y_t
\nonumber  \,\, ,
\\ 
G_{2}(y_t)&=& \frac{y_t(y_t-3)}{4 (y_t-1)^2}+\frac{y_t} {2(y_t-1)^3} \, 
\ln y_t  \nonumber\,\, ,
\\
H_{1}(y_t)&=& \frac{1-4 sin^2\theta_W}{sin^2\theta_W}\,\, \frac{xy_t}{8}\,
\left[ \frac{1}{y_t-1}-\frac{1}{(y_t-1)^2} \ln y_t \right]\nonumber \\
&-&
y_t \left[\frac{47 y_t^2-79 y_t+38}{108 (y_t-1)^3}-
\frac{3 y_t^3-6 y_t+4}{18(y_t-1)^4} \ln y_t \right] 
\nonumber  \,\, , 
\\ 
L_{1}(y_t)&=& \frac{1}{sin^2\theta_W} \,\,\frac{x y_t}{8}\, 
\left[-\frac{1}{y_t-1}+ \frac{1}{(y_t-1)^2} \ln y_t \right]
\nonumber  \,\, .
\\ 
\label{F1G1}
\end{eqnarray}
Finally, the initial values of the coefficients in the model III are
\begin {eqnarray}   
C_i^{2HDM}(m_{W})&=&C_i^{SM}(m_{W})+C_i^{H}(m_{W}) , \nnb \\
C_{Q_{1}}^{2HDM}(m_{W})&=& \int^{1}_{0}dx \int^{1-x}_{0} dy \,
(C^{H^{0}}_{Q_{1}}((\bar{\xi}^{U}_{N,tt})^{2})+
 C^{H^{0}}_{Q_{1}}(\bar{\xi}^{U}_{N,tt})+
 C^{H^{0}}_{Q_{1}}(g^{4})+C^{h^{0}}_{Q_{1}}((\bar{\xi}^{U}_{N,tt})^{3}) \nnb
\\ & + &
 C^{h^{0}}_{Q_{1}}((\bar{\xi}^{U}_{N,tt})^{2})+
 C^{h^{0}}_{Q_{1}}(\bar{\xi}^{U}_{N,tt})+
 C^{h^{0}}_{Q_{1}}(\bar{\xi}^{D}_{N,bb})) , \nnb  \\
 C_{Q_{2}}^{2HDM}(m_{W})&=& \int^{1}_{0}dx \int^{1-x}_{0} dy\,
(C^{A^{0}}_{Q_{2}}((\bar{\xi}^{U}_{N,tt})^{3})+
C^{A^{0}}_{Q_{2}}((\bar{\xi}^{U}_{N,tt})^{2})+
 C^{A^{0}}_{Q_{2}}(\bar{\xi}^{U}_{N,tt})+
 C^{A^{0}}_{Q_{2}}(\bar{\xi}^{D}_{N,bb}))\nnb \\
C_{Q_{3}}^{2HDM}(m_W) & = & \frac{m_b}{m_{\tau} \sin^2 \theta_W} 
 (C_{Q_{1}}^{2HDM}(m_W)+C_{Q_{2}}^{2HDM}(m_W)) \nnb \\
C_{Q_{4}}^{2HDM}(m_W) & = & \frac{m_b}{m_{\tau} \sin^2 \theta_W} 
 (C_{Q_{1}}^{2HDM}(m_W)-C_{Q_{2}}^{2HDM}(m_W)) \nnb \\
C_{Q_{i}}^{2HDM} (m_W) & = & 0\,\, , \,\, i=5,..., 10.
\label{CiW}
\end{eqnarray}
Here, we present $C_{Q_{1}}$ and $C_{Q_{2}}$ in terms of the Feynmann
parameters $x$ and $y$ since the integrated results are extremely large.
Using these initial values, we can calculate the coefficients 
$C_{i}^{2HDM}(\mu)$ and $C^{2HDM}_{Q_i}(\mu)$ 
at any lower scale in the effective theory 
with five quarks, namely $u,c,d,s,b$ similar to the SM case 
\cite{Buras,Chao,Alil2,Misiak}. 

The Wilson  coefficients playing  the essential role 
in this process are $C_{7}^{2HDM}(\mu)$, $C_{9}^{2HDM}(\mu)$,
$C_{10}^{2HDM}(\mu)$, 
$C^{2HDM}_{Q_1}(\mu )$ and $C^{2HDM}_{Q_2}(\mu )$. For completeness,
in the following we give their explicit expressions. 
\begin{eqnarray}
C_{7}^{eff}(\mu)&=&C_{7}^{2HDM}(\mu)+ Q_d \, 
(C_{5}^{2HDM}(\mu) + N_c \, C_{6}^{2HDM}(\mu))\nonumber \, \, ,
\label{C7eff}
\end{eqnarray}
where the LO  QCD corrected Wilson coefficient 
$C_{7}^{LO, 2HDM}(\mu)$ is given by
\begin{eqnarray} 
C_{7}^{LO, 2HDM}(\mu)&=& \eta^{16/23} C_{7}^{2HDM}(m_{W})+(8/3) 
(\eta^{14/23}-\eta^{16/23}) C_{8}^{2HDM}(m_{W})\nonumber \,\, \\
&+& C_{2}^{2HDM}(m_{W}) \sum_{i=1}^{8} h_{i} \eta^{a_{i}} \,\, , 
\label{LOwils}
\end{eqnarray}
and $\eta =\alpha_{s}(m_{W})/\alpha_{s}(\mu)$, $h_{i}$ and $a_{i}$ are 
the numbers which appear during the evaluation \cite{Buras}. 

$C_9^{eff}(\mu)$ contains a perturbative part and a part coming from LD
effects due to conversion of the real $\bar{c}c$ into lepton pair $\tau^+
\tau^-$:
\begin{eqnarray}
C_9^{eff}(\mu)=C_9^{pert}(\mu)+ Y_{reson}(s)\,\, ,
\label{C9efftot}
\end{eqnarray}
where
\newpage
\begin{eqnarray} 
C_9^{pert}(\mu)&=& C_9^{2HDM}(\mu) \nonumber 
\\ &+& h(z,  s) \left( 3 C_1(\mu) + C_2(\mu) + 3 C_3(\mu) + 
C_4(\mu) + 3 C_5(\mu) + C_6(\mu) \right) \nonumber \\
&- & \frac{1}{2} h(1, s) \left( 4 C_3(\mu) + 4 C_4(\mu) + 3
C_5(\mu) + C_6(\mu) \right) \\
&- &  \frac{1}{2} h(0,  s) \left( C_3(\mu) + 3 C_4(\mu) \right) +
\frac{2}{9} \left( 3 C_3(\mu) + C_4(\mu) + 3 C_5(\mu) + C_6(\mu)
\right) \nonumber \,\, ,
\label{C9eff2}
\end{eqnarray}
and
\begin{eqnarray}
Y_{reson}(s)&=&-\frac{3}{\alpha^2_{em}}\kappa \sum_{V_i=\psi_i}
\frac{\pi \Gamma(V_i\rightarrow \tau^+ \tau^-)m_{V_i}}{q^2-m_{V_i}+i m_{V_i}
\Gamma_{V_i}} \nonumber \\
& & \left( 3 C_1(\mu) + C_2(\mu) + 3 C_3(\mu) + 
C_4(\mu) + 3 C_5(\mu) + C_6(\mu) \right).
\label{Yres}
\end{eqnarray}
In eq.(\ref{C9efftot}), the functions $h(u, s)$ are given by
\begin{eqnarray}
h(u, s) &=& -\frac{8}{9}\ln\frac{m_b}{\mu} - \frac{8}{9}\ln u +
\frac{8}{27} + \frac{4}{9} x \\
& & - \frac{2}{9} (2+x) |1-x|^{1/2} \left\{\begin{array}{ll}
\left( \ln\left| \frac{\sqrt{1-x} + 1}{\sqrt{1-x} - 1}\right| - 
i\pi \right), &\mbox{for } x \equiv \frac{4u^2}{ s} < 1 \nonumber \\
2 \arctan \frac{1}{\sqrt{x-1}}, & \mbox{for } x \equiv \frac
{4u^2}{ s} > 1,
\end{array}
\right. \\
h(0,s) &=& \frac{8}{27} -\frac{8}{9} \ln\frac{m_b}{\mu} - 
\frac{4}{9} \ln s + \frac{4}{9} i\pi \,\, , 
\label{hfunc}
\end{eqnarray}
with $u=\frac{m_c}{m_b}$.
The phenomenological parameter $\kappa$ in eq. (\ref{Yres}) is taken as 
$2.3$. In eqs. (30) and (\ref{Yres}), the contributions of 
the coefficients $C_1(\mu)$, ...., $C_6(\mu)$ are due to the operator mixing.

Finally, the Wilson coefficients $C_{Q_1}(\mu)$ and $C_{Q_2}(\mu )$  
are given by \cite{Dai}
\beq
C_{Q_i}(\mu )=\eta^{-12/23}\,C_{Q_i}(m_W)~,~i=1,2~. 
\eeq
\section{The form factors for the decay $B\rightarrow K^* l^+ l^-$ }
The structure functions appearing in eq. (\ref{matr2}) are 
\begin{eqnarray}
A &=& -C_9^{eff} \frac{V}{m_B + m_{K^*}} - 4 C_7^{eff} \frac{m_b}{q^2} T_1~,
\nonumber\\ 
B_1 &=& -C_9^{eff} (m_B + m_{K^*}) A_1 - 4 C_7^{eff} \frac{m_b}{q^2} (m_B^2 -
m_{K^*}^2) T_2~,  \nonumber \\ 
B_2 &=& -C_9^{eff} \frac{A_2}{m_B + m_{K^*}} - 4 C_7^{eff} \frac{m_b}{q^2} 
\ga T_2 + \frac{q^2}{m_B^2 - m_{K^*}^2} T_3 \dr~,  \nonumber \\
B_3 &=& -C_9^{eff}\frac{ 2 m_{K^*}}{  q^2}(A_3 - A_0) + 4 C_7^{eff} 
\frac{m_b}{q^2}T_3~,  \nonumber \\ 
C &=& -C_{10} \frac{V}{m_B + m_{K^*}}~,  \nonumber \\  
D_1 &=& -C_{10} (m_B + m_{K^*}) A_1~,  \nonumber \\     
D_2 &=& -C_{10} \frac{A_2}{m_B + m_{K^*}}~,  \nonumber \\ 
D_3 &=& -C_{10} \frac{2 m_{K^*}}{q^2} (A_3 - A_0)~, \nonumber  \\
F &=& C_{Q_1} \frac{2 m_{K^*}}{m_b} A_0~, \nonumber \\ 
G &=& C_{Q_2} \frac{2 m_{K^*}}{m_b} A_0~, 
\label{hadpar1}
\end{eqnarray}

We use the $q^2$ dependent expression which is calculated in the framework 
of light-cone QCD sum rules in \cite{Pall} to calculate the hadronic 
form factors $V,~A_1,~A_2,~A_0,~T_1,~T_2$ and $T_3$: 
\begin{eqnarray}
F(q^2)=\frac{F(0)}{1-a_F \frac{q^2}{m_B^2}+b_F (\frac{q^2}{m_B^2})^2}\, ,
\label{formfac}
\end{eqnarray}
where the values of parameters $F(0)$, $a_F$ and $b_F$ are listed in 
\ref{Table2}.

\begin{table}[h]
    \begin{center}
    \begin{tabular}{|c|c|c|c|}
    \hline
   % \multicolumn{1}{|c|}{Parameter} & 
   %             \multicolumn{1}{|c|}{Value}     \\
    \hline \hline
                &$F(0)$              &       $a_F$ &  $b_F$\\
    \hline \hline    
    $A_1$       &$0.34\pm 0.05$      &       $0.60$&  $-0.023$ \\
    $A_2$       &$0.28\pm 0.04$      &       $1.18$&  $ 0.281$ \\
    $V  $       &$0.46\pm 0.07$      &       $1.55$&  $ 0.575$ \\
    $T_1$       &$0.19\pm 0.03$      &       $1.59$&  $ 0.615$ \\
    $T_2$       &$0.19\pm 0.03$      &       $0.49$&  $-0.241$ \\
    $T_3$       &$0.13\pm 0.02$      &       $1.20$&  $ 0.098$ \\
    \hline
        \end{tabular}
        \end{center}
\caption{The values of parameters existing in eq.(\ref{formfac}) for 
the various form factors of the transition $B\rightarrow K^*$.} 
\label{Table2}
\end{table}
\end{appendix}
\newpage

\end{document}